\pdfoutput=1
\documentclass[apj]{emulateapj}

\usepackage{epstopdf}
\usepackage{natbib}
\usepackage{apjfonts}
\usepackage{amsmath}
\usepackage[caption=false]{subfig}
\usepackage[mediumspace,Gray,squaren]{SIunits}

\newcommand{\arone}{148\,GHz}
\newcommand{\artwo}{218\,GHz}
\newcommand{\arthree}{277\,GHz}

\newcommand{\commentx}[1]{}

\newcommand{\mat}[1]{\ensuremath{\mathbf #1}}   

\newcommand{\ra}[3]   
   {\makebox[1.5em][r]{#1}\makebox[1.5em][r]{#2} \makebox[2em][r]{#3}}

\newcommand{\hms}[3]  
   {${#1}^{\mathrm{h}}{#2}^{\mathrm{m}}{#3}^{\mathrm{s}}$}

\newcommand{\hmin}[2]  
   {\ensuremath{{#1}^{\mathrm{h}}{#2}^{\mathrm{m}}}}
   
\newcommand{\hours}[1]  
   {\ensuremath{{#1}^{\mathrm{h}}}}

\newcommand{\dms}[3]  
   {\ensuremath{{#1}\degree{#2}\arcminute{#3}\arcsecond}}

\newcommand{\dm}[2]  
   {\ensuremath{{#1}\degree{#2}\arcminute}}

\newcommand{\ukcmb}  
           {\ensuremath{\micro \kelvin_\mathrm{cmb}}}

\newcommand{\uk}  
           {\ensuremath{\micro \kelvin}}

\newcommand{\fdeg} 
           {\hbox{$.\!\!^{\circ}$}}

\usepackage{color}

\shorttitle{Data and Map Making}
\shortauthors{ACT Collaboration}
\submitted{Submitted to ApJS}

\begin{document}

\title{THE ATACAMA COSMOLOGY TELESCOPE: DATA CHARACTERIZATION AND MAP MAKING}
\author{
Rolando~D\"{u}nner\altaffilmark{1},
Matthew~Hasselfield\altaffilmark{2},
Tobias~A.~Marriage\altaffilmark{3,4},
Jon~Sievers\altaffilmark{5,6},
Viviana~Acquaviva\altaffilmark{3,7},
Graeme~E.~Addison\altaffilmark{8},
Peter~A.~R.~Ade\altaffilmark{9},
Paula~Aguirre\altaffilmark{1},
Mandana~Amiri\altaffilmark{2},
John~William~Appel\altaffilmark{5},
L.~Felipe~Barrientos\altaffilmark{1},
Elia~S.~Battistelli\altaffilmark{10,2},
J.~Richard~Bond\altaffilmark{6},
Ben~Brown\altaffilmark{11},
Bryce~Burger\altaffilmark{2},
Erminia~Calabrese\altaffilmark{8},
Jay~Chervenak\altaffilmark{12},
Sudeep~Das\altaffilmark{13,3,5},
Mark~J.~Devlin\altaffilmark{14},
Simon~R.~Dicker\altaffilmark{14},
W.~Bertrand~Doriese\altaffilmark{15},
Joanna~Dunkley\altaffilmark{5,8,3},
Thomas~Essinger-Hileman\altaffilmark{5},
Ryan~P.~Fisher\altaffilmark{5},
Megan~B.~Gralla\altaffilmark{4},
Joseph~W.~Fowler\altaffilmark{15,5},
Amir~Hajian\altaffilmark{16,3,5},
Mark~Halpern\altaffilmark{2},
Carlos~Hern\'andez-Monteagudo\altaffilmark{17},
Gene~C.~Hilton\altaffilmark{15},
Matt~Hilton\altaffilmark{18,19},
Adam~D.~Hincks\altaffilmark{5,6},
Ren\'ee~Hlozek\altaffilmark{8,3},
Kevin~M.~Huffenberger\altaffilmark{20},
David~H.~Hughes\altaffilmark{21},
John~P.~Hughes\altaffilmark{7},
Leopoldo~Infante\altaffilmark{1},
Kent~D.~Irwin\altaffilmark{15},
Jean~Baptiste~Juin\altaffilmark{1},
Madhuri~Kaul\altaffilmark{14},
Jeff~Klein\altaffilmark{14},
Arthur~Kosowsky\altaffilmark{11},
Judy~M~Lau\altaffilmark{5},
Michele~Limon\altaffilmark{22,14,5},
Yen-Ting~Lin\altaffilmark{23,3,1,24},
Thibaut~Louis\altaffilmark{8},
Robert~H.~Lupton\altaffilmark{3},
Danica~Marsden\altaffilmark{25},
Krista~Martocci\altaffilmark{26,5},
Phil~Mauskopf\altaffilmark{9},
Felipe~Menanteau\altaffilmark{7},
Kavilan~Moodley\altaffilmark{18,27},
Harvey~Moseley\altaffilmark{12},
Calvin~B~Netterfield\altaffilmark{16},
Michael~D.~Niemack\altaffilmark{15,5},
Michael~R.~Nolta\altaffilmark{6},
Lyman~A.~Page\altaffilmark{5},
Lucas~Parker\altaffilmark{5},
Bruce~Partridge\altaffilmark{28},
Hernan~Quintana\altaffilmark{1},
Beth~Reid\altaffilmark{13,5},
Neelima~Sehgal\altaffilmark{3},
Blake~D.~Sherwin\altaffilmark{5},
David~N.~Spergel\altaffilmark{3},
Suzanne~T.~Staggs\altaffilmark{5},
Daniel~S.~Swetz\altaffilmark{14,15},
Eric~R.~Switzer\altaffilmark{6,5},
Robert~Thornton\altaffilmark{14,29},
Hy~Trac\altaffilmark{30,3},
Carole~Tucker\altaffilmark{9},
Ryan~Warne\altaffilmark{18},
Grant~Wilson\altaffilmark{31},
Ed~Wollack\altaffilmark{12},
Yue~Zhao\altaffilmark{5}
}
\altaffiltext{1}{Departamento de Astronom{\'{i}}a y Astrof{\'{i}}sica, Facultad de F{\'{i}}sica, Pontificia Universidad Cat\'{o}lica de Chile, Casilla 306, Santiago 22, Chile}
\altaffiltext{2}{Department of Physics and Astronomy, University of British Columbia, Vancouver, BC V6T 1Z4, Canada}
\altaffiltext{3}{Department of Astrophysical Sciences, Peyton Hall, Princeton University, Princeton, NJ 08544, USA}
\altaffiltext{4}{Department of Physics and Astronomy, The Johns Hopkins University, 3400 N. Charles St., Baltimore, MD 21218-2686, USA}
\altaffiltext{5}{Joseph Henry Laboratories of Physics, Jadwin Hall, Princeton University, Princeton, NJ 08544, USA}
\altaffiltext{6}{Canadian Institute for Theoretical Astrophysics, University of Toronto, Toronto, ON M5S 3H8, Canada}
\altaffiltext{7}{Department of Physics and Astronomy, Rutgers, The State University of New Jersey, Piscataway, NJ 08854-8019, USA}
\altaffiltext{8}{Department of Astrophysics, Oxford University, Oxford OX1 3RH, UK}
\altaffiltext{9}{School of Physics and Astronomy, Cardiff University, The Parade, Cardiff, Wales CF24 3AA, UK}
\altaffiltext{10}{Department of Physics, University of Rome ``La Sapienza'', Piazzale Aldo Moro 5, I-00185 Rome, Italy}
\altaffiltext{11}{Department of Physics and Astronomy, University of Pittsburgh, Pittsburgh, PA 15260, USA}
\altaffiltext{12}{Code 553/665, NASA/Goddard Space Flight Center, Greenbelt, MD 20771, USA}
\altaffiltext{13}{Berkeley Center for Cosmological Physics, LBL and Department of Physics, University of California, Berkeley, CA 94720, USA}
\altaffiltext{14}{Department of Physics and Astronomy, University of Pennsylvania, 209 South 33rd Street, Philadelphia, PA 19104, USA}
\altaffiltext{15}{NIST Quantum Devices Group, 325 Broadway Mailcode 817.03, Boulder, CO 80305, USA}
\altaffiltext{16}{Department of Physics, University of Toronto, 60 St. George Street, Toronto, ON M5S 1A7, Canada}
\altaffiltext{17}{Max Planck Institut f\"ur Astrophysik, Postfach 1317, D-85741 Garching bei M\"unchen, Germany}
\altaffiltext{18}{Astrophysics and Cosmology Research Unit, School of Mathematical Sciences, University of KwaZulu-Natal, Durban, 4041, South Africa}
\altaffiltext{19}{Centre for Astronomy \& Particle Theory, School of Physics \& Astronomy, University of Nottingham, Nottingham NG7 2RD, UK}
\altaffiltext{20}{Department of Physics, University of Miami, Coral Gables, FL 33124, USA}
\altaffiltext{21}{Instituto Nacional de Astrof\'isica, \'Optica y Electr\'onica (INAOE), Tonantzintla, Puebla, Mexico}
\altaffiltext{22}{Columbia Astrophysics Laboratory, 550 W. 120th St. Mail Code 5247, New York, NY 10027, USA}
\altaffiltext{23}{Institute for the Physics and Mathematics of the Universe, The University of Tokyo, Kashiwa, Chiba 277-8568, Japan}
\altaffiltext{24}{Institute of Astronomy \& Astrophysics, Academia Sinica, Taipei, Taiwan}
\altaffiltext{25}{Department of Physics, University of California Santa Barbara, CA 93106, USA}

\altaffiltext{26}{Kavli Institute for Cosmological Physics, Laboratory for Astrophysics and Space Research, 5620 South Ellis Ave., Chicago, IL 60637, USA}
\altaffiltext{27}{Centre for High Performance Computing, CSIR Campus, 15 Lower Hope St. Rosebank, Cape Town, South Africa}
\altaffiltext{28}{Department of Physics and Astronomy, Haverford College, Haverford, PA 19041, USA}
\altaffiltext{29}{Department of Physics , West Chester University of Pennsylvania, West Chester, PA 19383, USA}
\altaffiltext{30}{Harvard-Smithsonian Center for Astrophysics, Harvard University, Cambridge, MA 02138, USA}
\altaffiltext{31}{Department of Astronomy, University of Massachusetts, Amherst, MA 01003, USA}

\begin{abstract}
We present a description of the data reduction and mapmaking pipeline used for the 2008 observing season of the Atacama Cosmology Telescope (ACT).
The data presented here at 148 GHz represent 12\% of the 90\,TB collected by ACT from 2007 to 2010.
In 2008 we observed for 136 days, producing a total of 1423\,\hour~of data (11\,TB for the \arone~band only), with a daily average of 10.5\,\hour~of observation.
From these, 1085\,\hour~were devoted to a 850\,$\deg^2$ stripe (11.2\,\hour~by 9\fdeg1) centered on a declination of -52\fdeg7, while 175\,\hour~were devoted to a 280\,$\deg^2$ stripe (4.5\,\hour~by 4\fdeg8) centered at the celestial equator.
We discuss sources of statistical and systematic noise, calibration, telescope pointing, and data selection.
Out of 1260 survey hours and 1024 detectors per array, 816\,\hour~and 593 effective detectors remain after data selection for this frequency band, yielding a 38\% survey efficiency.
The total sensitivity in 2008, determined from the noise level between 5\,\hertz~and 20\,\hertz~in the time-ordered data stream (TOD), is $32\,\mu\kelvin\sqrt{\second}$ in CMB units. 
Atmospheric brightness fluctuations constitute the main contaminant in the data and dominate the detector noise covariance at low frequencies in the TOD.
The maps were made by solving the least-squares problem using the Preconditioned Conjugate Gradient method, incorporating the details of the detector and noise correlations.
Cross-correlation with WMAP sky maps, as well as analysis from simulations, reveal that our maps are unbiased at multipoles $\ell > 300$. 
This paper accompanies the public release of the \arone~southern stripe maps from 2008.
The techniques described here will be applied to future maps and data releases.

\end{abstract}

\maketitle

\setlength{\textfloatsep}{5mm}

\section{Introduction}
\label{sec_introduction}
Over the past two decades, precision measurements of the Cosmic Microwave Background (CMB) have led to remarkable advances in our understanding of cosmology.
Combined with other observations, they have produced constraints on models of the Universe to percent level accuracy (e.g., \citet{komatsu/etal:2011}). 
Primary CMB anisotropy has been measured to cosmic variance precision by WMAP up to
multipoles of  approximately 500 \citep{larson/etal:2011}. 
Observations at finer angular scales \citep{friedman/etal:2009, reichardt/etal:2009, reichardt/etal:2009a, veneziani/etal:2009, kessler/etal:2009, sievers/etal:prep, sharp/etal:2010, shirokoff/etal:2011, das/etal:2011, reichardt/etal:2011}, corresponding to the damping scale of the anisotropies, have led to even tighter parameter constraints, which are improving with each new data set.
Additionally, CMB measurements at smaller angular scales are sensitive to the contribution of point sources, which help to reveal the nature of early galaxies or active galactic nuclei \citep{vieira/etal:2010, marriage/etal:2011a}, to the thermal Sunyaev-Zel'dovich (SZ) effect, which can provide independent constraints on cosmological parameters \citep{sunyaev/zeldovich:1970,  marriage/etal:2011b, sehgal/etal:2011, planck/etal:2011, reichardt/etal:prep}, and to gravitational lensing effects on the CMB \citep{seljak:1996, zaldariaga/seljak:1999, das/etal:2011b, sherwin/etal:2011}.

The Atacama Cosmology Telescope (ACT) is located at \dms{22}{57}{35}S, \dms{67}{47}{13}W on Cerro Toco at an altitude of 5200\,\meter~in the Atacama Desert of northern Chile. 
Its main purposes are to map the millimeter wave sky at arcminute scales, sampling multipoles up to $l\simeq10^4$ and detecting and characterizing foregrounds, including galaxy clusters through their SZ signature, and millimeter galaxies.

Between 2007 and 2011, ACT was equipped with the Millimeter Bolometric Array Camera (MBAC), which observed simultaneously in three bands: \arone, \artwo~and \arthree. 
The bands were chosen to avoid major atmospheric emission lines and to sample the SZ decrement, null and increment.
Each band had a dedicated set of optics and a detector array composed of 1024 pop-up TES bolometers \citep{benford/etal:2003}  coupled to the optical signal, called ``live detectors,'' plus 32 ``dark detectors,'' which were not coupled to the sky.
For details about the instrument see \citet{swetz/etal:2011}.

Since first light on October 23, 2007, the telescope has had four observing seasons producing more than 90\,TB of data.  
Here we describe the 11\,TB of data taken at \arone~in 2008, including the techniques used for data characterization, as well as the data reduction process used to obtain the final maps. 
The \artwo~and \arthree~data will be described in a later paper.
However, a similar data reduction pipeline is used to analyze the full dataset.

The maps obtained from these data have an angular resolution of 1.37\arcminute~\citep{hincks/etal:2010} and a noise level that ranges between 25 and 50\,$\mu$\kelvin-arcmin. 
The calibration of the data to WMAP is discussed in \cite{hajian/etal:2011}.
The power spectrum of the maps is presented in \cite{fowler/etal:2010} and \cite{das/etal:2011}, with corresponding constraints on cosmological parameters in \cite{dunkley/etal:2011}. 
Cluster detections through their SZ signature are presented by \cite{hincks/etal:2010} and \cite{marriage/etal:2011b}, while extragalactic source detections are given in \cite{marriage/etal:2011a}. 
Multi-wavelength follow up for these clusters is presented in \cite{menanteau/etal:2010} as well as the discovery a massive cluster at $z=0.87$, \cite[El Gordo]{menanteau/etal:2012}. 
Their cosmological interpretation is discussed in \cite{sehgal/etal:2011}.
The first direct detection of gravitational lensing of the microwave background was made using these maps in \cite{das/etal:2011b}. This in turn demonstrates for the first time that microwave background data on their own favor cosmologies with an accelerating expansion \citep{sherwin/etal:2011}.

This paper is organized as follows.
A summary of observations is given in Section~\ref{observations}. 
In Section~\ref{signals} we provide background for understanding the sky signal as recorded in the time-ordered data stream (TOD). 
In Sections~\ref{atmosphere}, \ref{random_noise} and \ref{systematic_noise} respectively, we characterize the atmospheric, detector and systematic noise found in the TOD.
Data calibration into units of sky temperature fluctuations, $\delta\mathrm{T_{CMB}}$, is described in Section~\ref{calibration}.
Section~\ref{pointing} describes the pointing solution, while Section~\ref{time-constants} explains the detector time-constant determination method.
Data selection is described in Section~\ref{data_selection}, providing final statistics on the amount of data used for making the maps.
The mapmaking method is discussed in Section~\ref{map-making}.
We conclude in Section \ref{conclusions} and present a step-by-step summary of the data pipeline from raw data to maps. 
Appendix \ref{mode_removal} provides further details about finding and removing correlated modes from the data.

This paper accompanies the public release of the data through the NASA's LAMBDA site.\footnote{http://lambda.gsfc.nasa.gov/product/act.}

\section{Observations}
\label{observations}
ACT observes the sky by scanning the telescope in azimuth at a constant elevation of 50\fdeg5 as the sky moves across the field of view in time, resulting in a stripe-shaped observation area.
With this scan strategy, the instrument observes through a constant air mass, the cryogenics remain stable, the telescope's shape remains constant, and the local environment and instrumental offsets are sampled in a consistent way.
The time constants of the detectors, together with mechanical factors, limit the scan speed and turnaround acceleration.
Table~\ref{scan_parameters} gives a summary of the scan parameters used in all seasons.
The lower acceleration introduced in 2008 was needed to reduce vibrations in the optics (see more details in \S\ref{systematic_noise}). 

\begin{deluxetable}{rlll}
\tablecaption{Scan parameters}
\tablecolumns{4}
\tablewidth{230pt}
\tablehead{\colhead{Season} &
                     \colhead{2007} & 
                     \colhead{2008} &
                     \colhead{2009-2010} }
\startdata
	Elevation & 50\fdeg5 & 50\fdeg5 & 50\fdeg5 \\
	Scan Width & 9\fdeg6 & 7\fdeg0 & 7\fdeg0 \\
	Period & 19.4\,$\second$ & 10.2\,$\second$ & 10.2\,$\second$ \\
	Speed & 1.0\,$\mathrm{deg}/\second$ & 1.5\,$\mathrm{deg}/\second$ & 1.5\,$\mathrm{deg}/\second$ \\
	Max. Accel. & 8.1\,$\mathrm{deg}/\second^2$ & 3.3\,$\mathrm{deg}/\second^2$ & 3.3\,$\mathrm{deg}/\second^2$ \\
	Data file length & 15\,\minute & 15\,\minute & 10\,\minute \\ 
\enddata
\label{scan_parameters}
\end{deluxetable}%

The observations are repeated at complementary central azimuth angles to capture both rising and setting skies.  
This cross-linking technique helps minimize systematic effects due to the scan in the mapmaking process and improves the determination of CMB modes parallel to the scan direction.  

\begin{figure}
\begin{center}
\includegraphics[width=90mm]{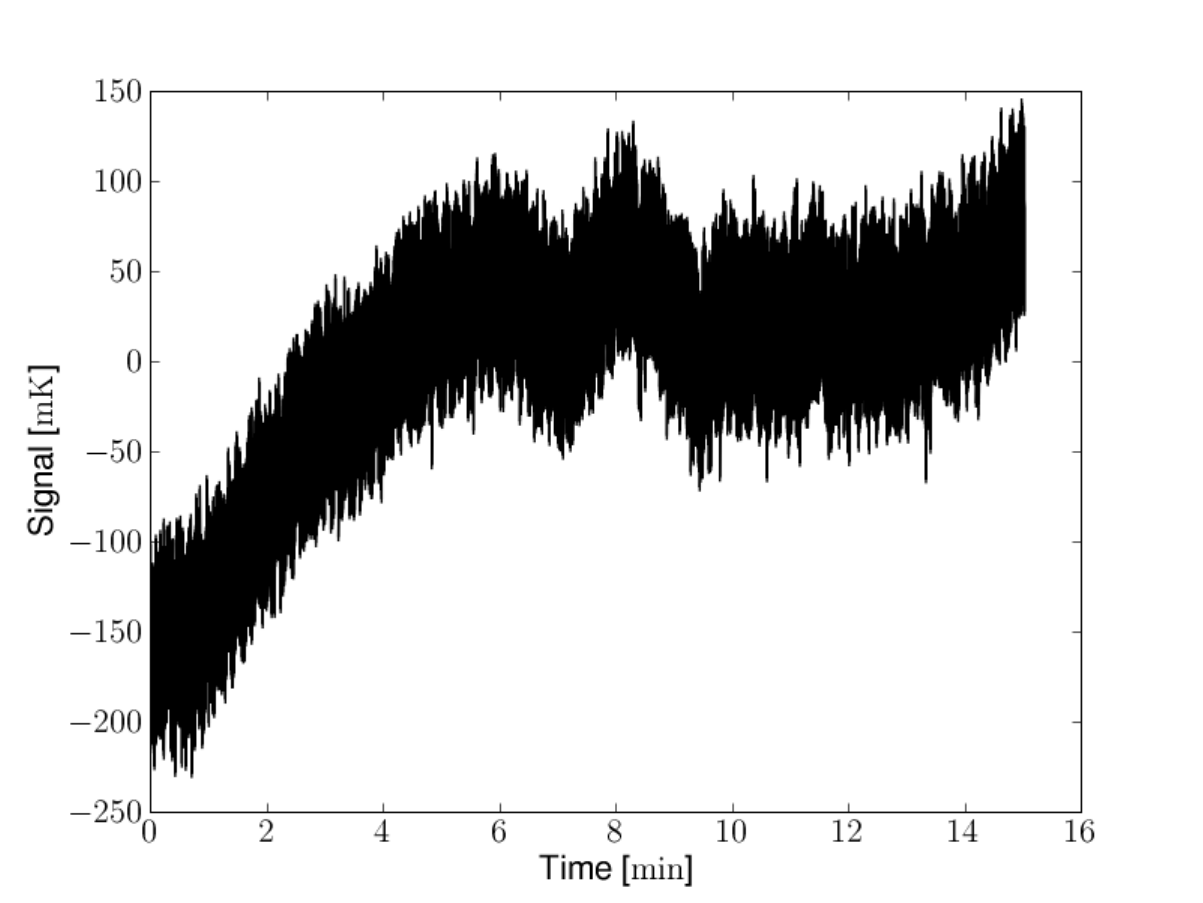}
\caption{\small{Example TOD from one detector at \arone~from October 21, 2008. 
This was a good observing night with a PWV of 0.22\,mm.  
The slow drift is dominated by changes in the atmosphere brightness.
The high-frequency noise is dominated by detector noise. Units are $\milli\kelvin$ in CMB equivalent units at \arone.
The plot displays $3.6\times10^5$ samples at an interval of $2.5\times10^{-3}$ seconds.
The telescope was scanning while these data were taken.}
}
\label{example_TOD_AR1}
\end{center}
\end{figure}

Each detector is sampled at a rate of 398.72\,\hertz~and the data are stored in 15-minute long data files.
They are then merged with the rest of the housekeeping data, which include the azimuth and elevation encoder readings and time of day. 
The data are compressed to one-third of their original size for storage, using a lossless compression algorithm (SLIM).\footnote{For further detail see \url{http://slimdata.sourceforge.net/}.}
A sample TOD is shown in Figure \ref{example_TOD_AR1} for a night with good weather conditions, meaning that precipitable water vapor (PWV) remained below 1\,\milli\meter. The PWV is a good indicator of \arone~opacity and overall data quality.

Despite some variation over the season due to changes in sunrise and sunset times, a typical day of observations at the site was as follows.
The cryogenic system was recycled every day before providing roughly 14 hours of observing time.
The cryogenic cycle began around 11:00 in the morning local time and lasted for 9.5 hours, so that MBAC was cold by about 20:30.  
At 20:40, warm-up movements for the motors and gears were run, and at 20:50 the detectors were biased and the first detector calibration data were obtained (see \S\ref{subsec:calTime}). 
The observations started at 21:00, usually by scanning the rising sky.
Around 2:30 more detector calibration data were collected and the scan was shifted to the west to observe the same region that was previously observed while rising.
Observing in the west has an additional advantage from a hardware safety perspective: in the event of a telescope failure, MBAC would be pointing away from the Sun at its rising.
Observations normally ended around 10:40, nearly 2 hours after sunrise above the mountains.\footnote{For pointing accuracy reasons, we ended up using only data obtained less than one hour after sunrise.}
At this point, final detector calibration data were taken, before the telescope was sent to its home position and the cycle was restarted.  
When the observable region of the sky contained a planet, it was normally scanned every other night for calibration and beam measurements.
This was not done every single night to avoid producing a gap in the CMB map.
All of the operations listed above were automated and could be performed remotely.

\begin{figure}
\begin{center}
\includegraphics[trim = 0 0 0 30, clip, width=90mm]{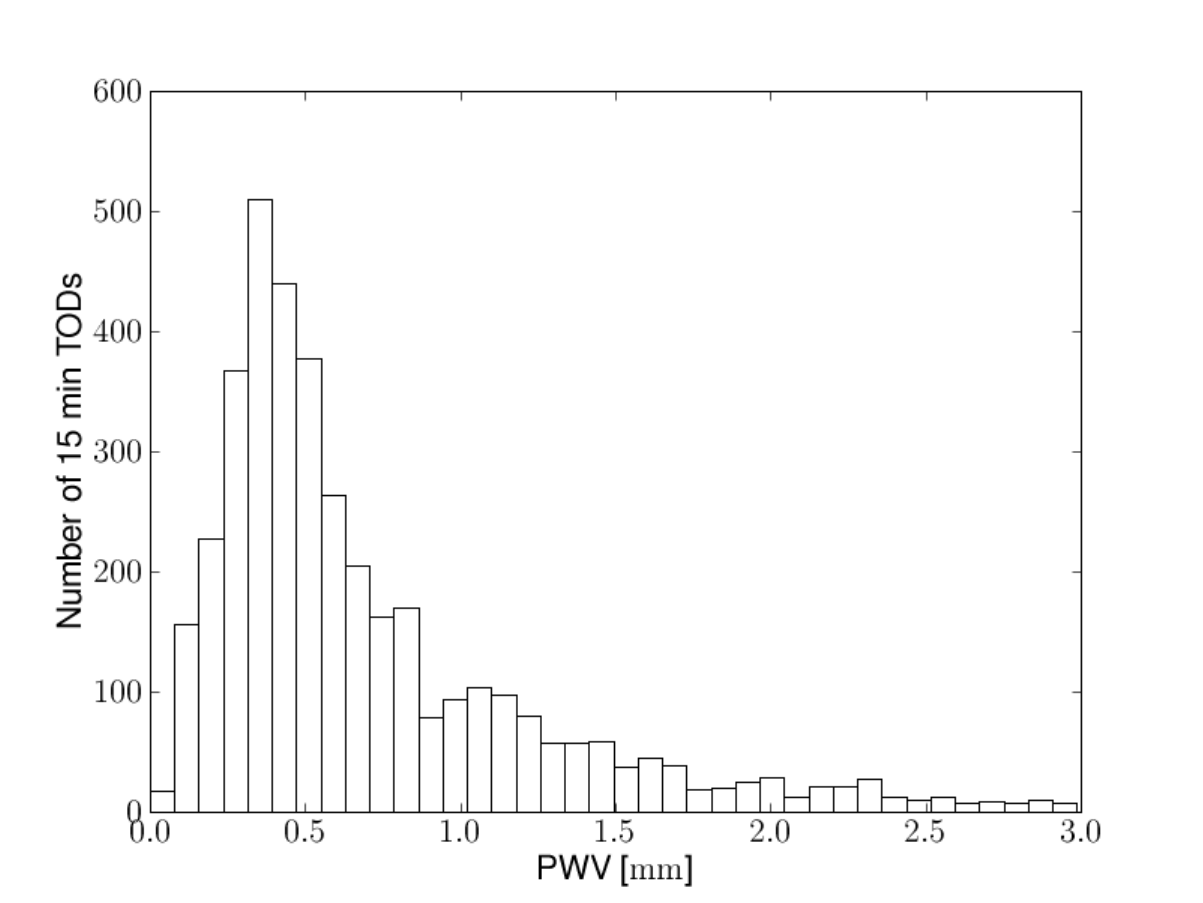}
\caption{\small{Histogram showing the PWV during the 2008 season, measured at the zenith of the Atacama Pathfinder Experiment (APEX) facility. 
The median value is 0.49~\milli\meter.}}
\label{pwv_season2008}
\end{center}
\end{figure}

The 2008 season began on August 11 and ended on December 24, with a total of 136 available nights and 124 nights with successful observations, resulting in 1423 hours and 25.7\,TB of total data for all three frequencies.  
From the total observed nights, four had bad weather conditions, leaving 120 nights with usable science observations.  
The overall calendar time efficiency, including day time, was 44\%.
The median PWV during observations across the season, measured at zenith of the Atacama Pathfinder Experiment (APEX) facility \citep{gusten/etal:2006}, was 0.49~\milli\meter.
Given that the ACT site is about 140\,\meter~higher than the APEX site, we use a correction factor $\mathrm{PWV}_{\mathrm{ACT}} = 0.88\,\mathrm{PWV}_{\mathrm{APEX}}$ to estimate the PWV at the ACT site.
Figure~\ref{pwv_season2008} shows a histogram of the PWV during the season.

Observations were made in two areas of the sky: The equatorial stripe, centered at a declination of 0\degree~(60\degree~and 300\degree~in azimuth at 50\fdeg5~elevation), and the southern stripe, centered at a declination of $-$53\degree~(150\degree and 210\degree azimuth at 50\fdeg5 elevation), covering a wide range in right ascension. 
Table~\ref{observation_summary} lists the boundaries of the observation areas and the number of hours available in each area before data selection.

\tabletypesize{\footnotesize} 
\begin{deluxetable}{ccccc}
\tablecolumns{5}
\tablecaption{Observation summary for seasons 2008 (at \arone).}
\tablewidth{0pt}
\tablehead{\colhead{Season} & 
                     \colhead{Dec (Min, Max)} & 
                     \colhead{RA (Min, Max)} & 
                     \colhead{Area $[\mathrm{deg}^2]$} & 
                     \colhead{Hours\tablenotemark{a}} }
\startdata
	\multicolumn{5}{l}{\em Southern Stripe}\\
	2008 & (-57\fdeg15, -48\fdeg1) & (\hmin{20}{43}, \hmin{7}{53}) & 850 & 1085 \\
	&&&&\\
	\multicolumn{5}{l}{\em Equatorial Stripe}\\
	2008 & (-2\fdeg12, 2\fdeg34) & (\hmin{10}{21}, \hmin{14}{48}) & 280 & 175 \\
\enddata
\tablenotetext{a}{Total hours before data selection.}
\label{observation_summary}
\end{deluxetable}%
\tabletypesize{\normalsize} 

\section{Signal Properties}
\label{signals}
In this section we characterize the expected TOD signal from point sources and extended sources, for comparison to the random and systematic noise described later in Sections \ref{random_noise} and \ref{systematic_noise}. 

\subsection{Point Sources}
For our purposes, we will define a point source as any object with an angular size comparable to or smaller than the beam size of the telescope.  
The beam full-width at half maximum (FWHM), $\theta_{1/2}$, is 1\farcm37 for \arone. 
Planets are the most important point sources as their high surface brightness makes them useful for both calibration and beam measurements.
Distant galaxies can also be approximated as point sources and are helpful for the pointing solution.

\begin{figure}[htbp]
  \begin{center}
    \includegraphics[width=90mm]{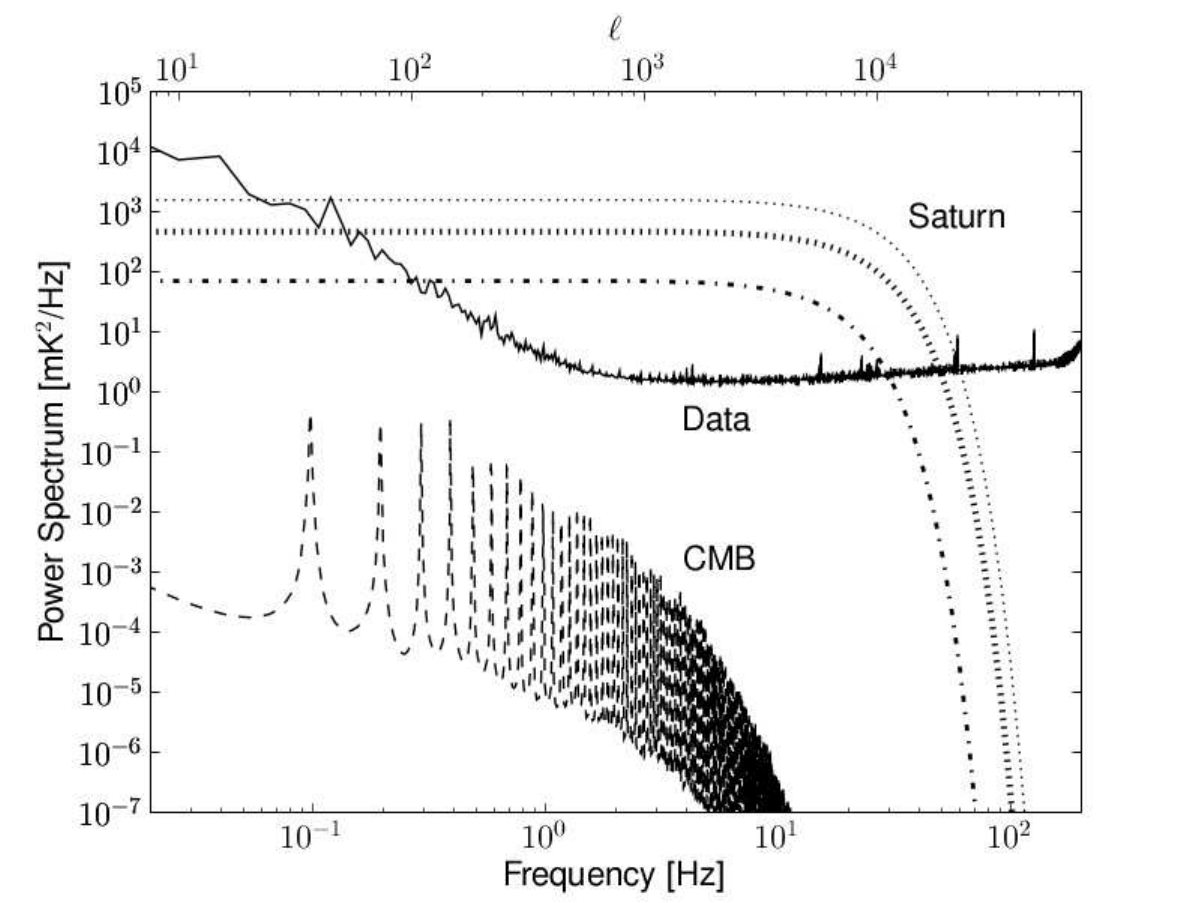}
    \caption{\small{Power spectra of signals in the 2008 season compared to the data.   
    The solid curve is the average power spectrum from 737 live detector TODs from one 15-minute 148 GHz data file, during which the PWV was 0.5 mm.
    The rise at low frequencies is the noise contribution from the atmosphere.
    The dashed curve shows the simulated CMB signal.
    The oscillations in the simulated CMB signal are due to enhanced power at harmonics of the scan frequency.
    The dot-dashed curve estimates the point source contribution (considering only one hit at the beam center) assuming a Gaussian beam.
    For comparison, the thin and thick dotted curves show the expected response for \artwo~and \arthree, with beam sizes of 1\farcm00 and 0\farcm91 respectively.  
    The amplitude of the point source power spectra corresponds to approximately the signal of Saturn.
    }} 
\label{signal_ps_2008}
\end{center}
\end{figure}

As the telescope scans and the sky rotates, a single detector traces a zigzag on the sky, sampling the point source every time it intersects with the beam. 
The number of times a point source appears depends on the scan period, the beam size and the central scan azimuth.
A point source appears as a succession of  "blips" in the time stream.
The shape of these "blips" is a slice through the telescope point spread function and depends upon the angular separation between the center of the beam and the location of the source.
The angular speed of the scan on the sky is given by $\dot{\theta} = v_{\mathrm{scan}}\cos{(50\fdeg5)}$.
The 2008 scan speed of $v_{\mathrm{scan}} = 1\fdeg5/\second$ implies an angular sky speed of $57\farcm2/\second$. 
Assuming a Gaussian-like beam of equivalent width and neglecting the transit speed of the source, the 3 dB cutoff frequency is
\begin{equation}
f_{-3\deci\bel} = \frac{2\ln{(2)}\>\dot{\theta}}{\pi \theta_{1/2}}.
\end{equation}
This means that the contribution from point sources to the TOD is limited to frequencies below $18\,\hertz$, as can be seen in Figure \ref{signal_ps_2008}.

Given our scan speed, nominal elevation, sky rotation and sampling rate, we expect nearly 10 samples per beam on the sky at \arone.

\subsection{Extended Sources}

Using CMB simulations we can estimate the expected TOD response to extended sky features.
Averaging the power spectra from many such synthetic TODs, we can estimate the CMB power spectrum in TOD space. 
Figure 3 shows the average power spectrum of a simulated 148 GHz observation of the CMB sky and the average data power spectrum from one 15-minute file.  
Because of Silk damping, the CMB has a relatively sharp cutoff in its characteristic size: unlike the point source observations shown in the figure, the CMB has little signal at frequencies higher than 10\,\hertz.

Given ACT's angular scan speed, the TOD frequency associated with a sky feature of angular size $\theta_f$ is
\begin{equation}
f_{\rm TOD} = \frac{\dot\theta}{2\theta_f}
\end{equation}
where the factor of 2 in the denominator comes from considering that the angular scale is the size of a positive or negative temperature bump, which is half of a wavelength on the sky.  
Given that multipole moment $\ell$ relates to angular scale as $\ell\simeq\pi/\theta_f$ for scales small compared to the full sky, an approximate conversion between multipole and TOD frequency is 
\begin{equation}
\ell\simeq \frac{2\pi f_{\rm TOD}}{\dot\theta} = \frac{2\pi f_{\rm TOD}}{v_{\mathrm{scan}}\cos{(50\fdeg5)}}.
\label{f_tod_vs_ell}
\end{equation}
Note that this relation is only a rule of thumb; the actual mapping of TOD frequency into multipole space depends on the specific scan strategy.  
This relation shows that the CMB power spectrum in the TOD can be shifted in frequency by changing the scan rate.
As a reference, the TOD frequency corresponding to $\ell=3000$ was 7.9\,\hertz~in the 2008 season.


\section{Atmospheric Emission}
\label{atmosphere}
The atmosphere is a strong emitter and absorber at the bands of interest, chiefly due to the excitation of the vibrational and rotational modes of water vapor. 
For this reason, the PWV is strongly correlated to the  level of optical loading on the detectors. 
The Atmospheric Transmission at Microwaves (ATM) model \citep{pardo/etal:2001} provides an estimate of the loading as a function of the PWV level. 
At the median value of 0.49\,\milli\meter~during the 2008 observations, the loading is approximately 0.5\,\pico\watt, which corresponds to an equivalent Rayleigh-Jeans temperature of 6.4\,\kelvin~at the nominal elevation. 

During the season, the median atmospheric temperature drift over 15-minute observations was $0.22\,\kelvin$ (Rayleigh Jeans equivalent units as measured by MBAC), with lower and upper quartiles at 0.10\,\kelvin~and 0.43\,\kelvin~respectively.

Turbulence induces a spatial structure in the atmospheric signal.
According to the Kolmogorov model of turbulence \citep{tatarskii:1961}, the power spectrum of the fluctuations in a large 3-dimensional volume is proportional to $q^{-11/3}$, where $q$ is the wavenumber.
The projected signal observed on the sky can have either a $q^{-11/3}$ dependence, if the wavelength is small enough that the turbulence can be treated as three dimensional, or a $q^{-8/3}$ dependence, if the wavelength is large compared to the thickness of the atmospheric layer supporting the turbulent motions \citep{church:1995,lay/halverson:2000}.

\begin{figure}[tbp]
\begin{center}
  \subfloat[]{\label{stare_atm_ps} 
  \includegraphics[trim = 20 0 15 0, clip, width = 85mm]{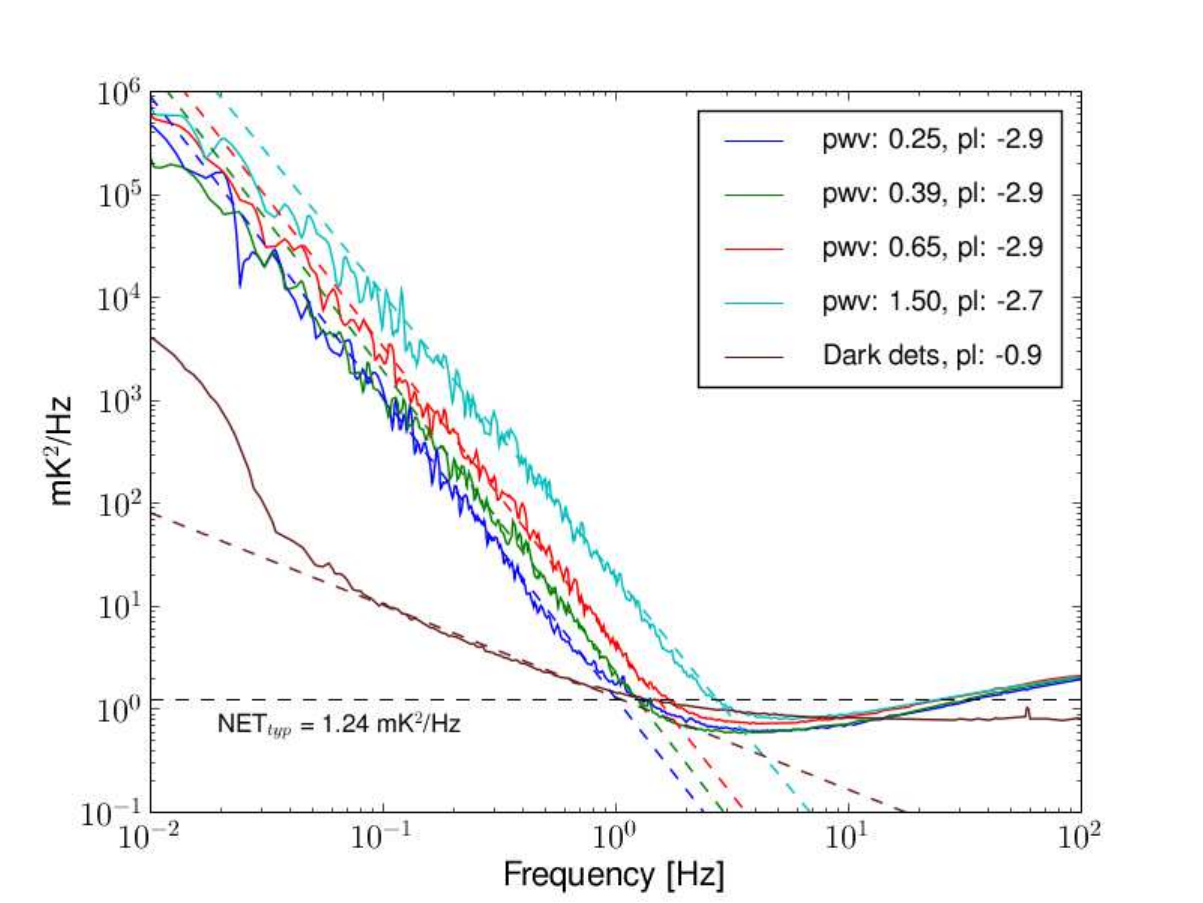}}
  \qquad
  \subfloat[]{\label{pl_vs_freq} 
  \includegraphics[trim = 20 0 15 0, clip, width = 85mm]{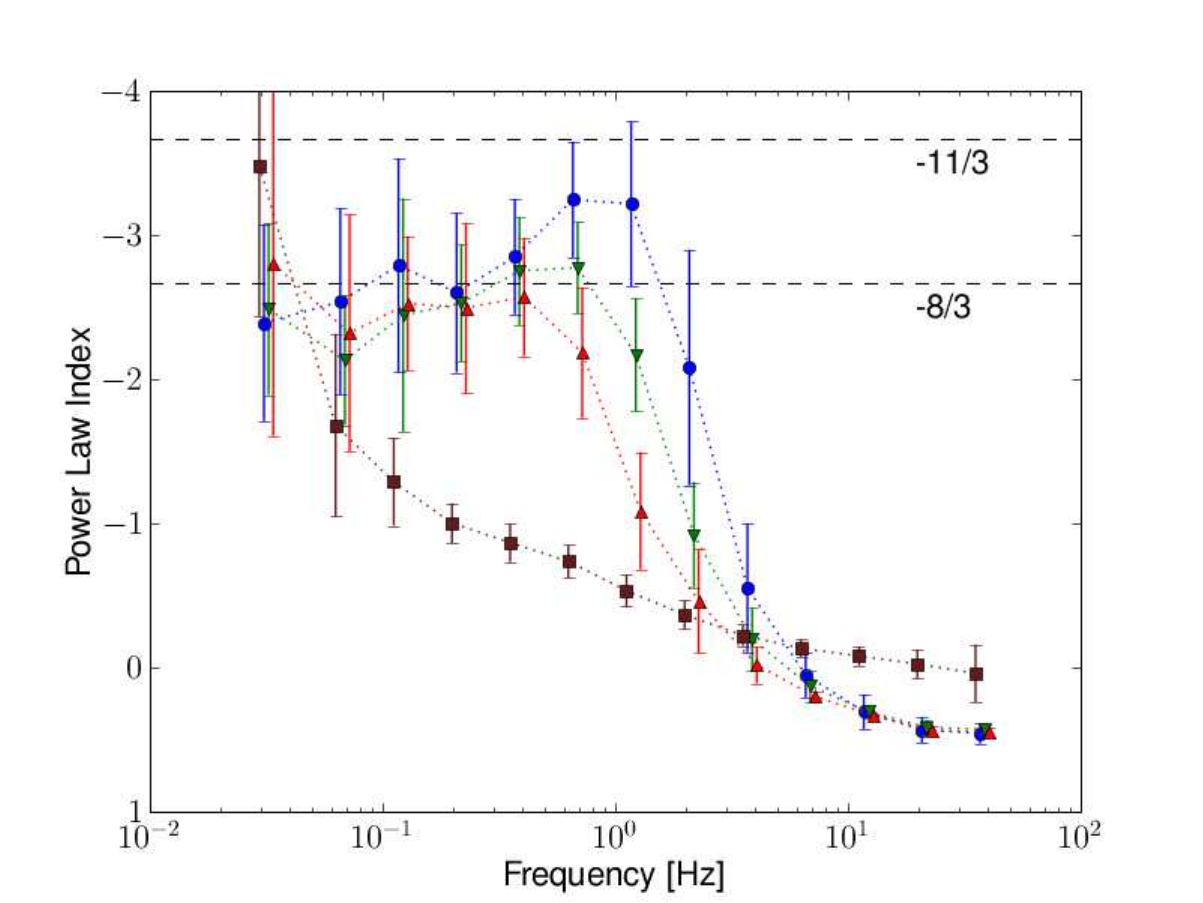}}
  \caption{\small{Atmospheric signature in TOD power spectrum. 
  (a) Average power spectra for four groups of non-scanning TODs selected by PWV level.
  The average power spectrum from the dark detectors was subtracted from each group power spectrum and is shown for comparison.
  The dashed lines show a power law fit to each spectrum.
  The legend indicates the mean PWV and the power-law index from the fit for each group.
  Logarithmic binning was used to reduce the noise in the plot.
  (b) Power law index as a function of frequency for the average power spectra from three groups (blue ($\bullet$), green ($\blacktriangledown$) and red ($\blacktriangle$)) of observations selected by their power law index at frequencies close to the knee, all belonging to the third PWV selected group in (a).
  Each value was obtained by fitting a power law to the corresponding power spectrum in a small frequency range (in logarithmic space); the listed frequencies are the mean values of the associated frequency range.
  The error bars show the dispersion of the indexes from the members of each group.
  The same is shown in brown (${\scriptstyle\blacksquare}$) for the dark detector average power spectrum for comparison.
   }} 
\label{stare_plot_AR1-r25c6}
\end{center}
\end{figure}

Figure~\ref{stare_plot_AR1-r25c6} shows examples of the atmospheric signature in TOD power spectra.
Figure~\ref{stare_atm_ps} displays the average TOD power spectra from four groups of observations with the telescope not scanning (``stare'' observations).
When scanning, the knee --where the power law meets the white noise level-- increases in frequency by around 1\,\hertz~and harmonics of the scan frequency leak into the spectrum.
Before averaging the power spectra from different observations (data files), the average power spectrum from the dark detectors was subtracted to isolate the atmospheric signal from instrumental $1/f$ noise.
TODs were binned as a function of the PWV level measured by APEX during the period of data acquisition.
It is clear how the atmosphere signal grows with PWV, shifting the knee towards higher frequencies.
On the other hand, the power law index stays rather constant near the 2-dimensional regime value.
The average power spectrum from the dark detectors is shown for comparison.
It is dominated by the thermal fluctuations of the cryostat.
The fact that the ``dark-detector'' power spectrum appears higher than the others in the frequency range between the knee and 10\,\hertz is the result of the power subtraction mentioned above.
This indicates that the $1/f$ plus readout noise dominates over the atmospheric plus detector noise in that frequency range.
The knee frequency ranges from 1 to 5\,\hertz~for stare observations depending on the weather conditions.

The departure from a pure power law shown in Fig.~\ref{stare_atm_ps} implies that the power law index varies with frequency.
By fitting power laws in small frequency ranges, shown as dashed lines in Fig.~\ref{stare_atm_ps}, we were able to measure this dependency and group the observations as a function of their power spectrum slopes at frequencies near the knee.
We found that, for similar PWV conditions, the slopes can vary significantly, with indices ranging between the 2-dimensional and 3-dimensional regimes, as shown in Figure \ref{pl_vs_freq}.
Low frequencies, which are related to large scales in the sky, are dominated by the 2-dimensional regime, with a power law index of approximately $\alpha =-2.4$.
At frequencies around 1\,\hertz~the power spectrum from some observations (blue dots in Fig. \ref{pl_vs_freq}) follow a steeper power law, suggesting that, under certain weather conditions, the 3-dimensional regime dominates at these smaller scales.
The slopes tend to increase with wind speed.
Higher wind speeds are expected to shift larger features of the turbulent pattern towards higher TOD frequencies.
As large scales are dominated by the 2-dimensional regime, this would cause the opposite effect of reducing the slopes at higher frequencies, in contradiction with what is observed.
This result suggests that higher winds might be associated with intrinsic properties of the turbulent layer, such as its width or height, producing a steeper power law \citep{church:1995,lay/halverson:2000}.

Atmospheric structures larger than 24\arcminute~--the field of view of a detector array-- appear as a common mode among detectors, while smaller features can produce sub-array ``correlated modes,'' as described in Appendix \ref{mode_removal}. 
Given that the distance to the turbulent layer is commonly less than a kilometer \citep{robson:2002}, it becomes out of focus, smearing out to scales of nearly $10\arcminute$, in agreement with our optical simulations.
This corresponds to roughly a third of the array size. 
Then, given the scan speed, such signals appear at frequencies between 2 or $3\,\hertz$ in the TOD. 
In general, the common mode of the detectors is a good estimator of the atmospheric signal, but the estimate can in principle be improved by dividing the array to account for sub-array atmosphere structure.
Our attempts to detect coherent motions of atmospheric features across the array, as would appear in a moving frozen-sheet model of the turbulent layer, were not successful. 
Instead, we suppress the atmospheric noise by solving for the strongest atmospheric modes in the time stream during the mapmaking process, as discussed in \S\ref{map-making}.

\section{System Sensitivity: Uncorrelated Noise}
\label{random_noise}

Above the atmosphere knee frequency, the TOD is dominated by broadband random noise.  
This noise is generated at the detectors and readout circuitry and it is essentially uncorrelated among detectors. 
After de-projecting the correlated modes from the atmosphere and systematics (see Section \ref{systematic_noise} and Appendix~\ref{mode_removal}), this noise can be measured by averaging the TOD power spectral density (PSD) over the desired range of frequencies.

The total sensitivity of the array, expressed as the noise equivalent temperature (NET), is given by

\begin{equation}
  \mathrm{NET}_{\mathrm{tot}} =
  \left[\sum_{i=0}^{N_{\mathrm{det}}}{\frac{1}{\mathrm{NET}_i^2}}\right]^{-1/2},
\end{equation}

\noindent where $\mathrm{NET}_i$ is the NET of each working detector. Thus, the typical sensitivity per detector can be defined as $\mathrm{NET}_{\mathrm{typ}} =  \mathrm{NET}_{\mathrm{tot}}\,\sqrt{N_{\mathrm{det}}}$, where $N_{\mathrm{det}}$ is the number of the ``effective'' detectors, as defined in \S\ref{data_selection}.

\tabletypesize{\footnotesize} 
\begin{deluxetable}{ccc}
\tablecolumns{3}
\tablewidth{230pt}
\scriptsize
\tablecaption{Median value of the typical sensitivity per detector and the total sensitivity for the array at two frequency ranges after removing the main 28 modes of correlated noise. The units are equivalent temperature for CMB fluctuations.}
\tablewidth{0pt}
\tablehead{\colhead{Range} & 
           \colhead{NET$_{\mathrm{typ}}$ [$\mu\kelvin\sqrt{\second}$]} &
           \colhead{NET$_{\mathrm{tot}}$ [$\mu\kelvin\sqrt{\second}$]}}
\startdata
5--20\,\hertz\tablenotemark{a} & $786\pm28$ &  $31.7\pm2.7$\\
100--120\,\hertz & $1052\pm21$ & $42.7\pm3.2$  \\
\enddata
\tablenotetext{a}{The signal band is $\leq30\,\hertz$ so these entries best estimate the instrument sensitivity.}
\label{sensitivities}
\end{deluxetable}%

Table~\ref{sensitivities} lists the typical values of the total and average NET for a mid-frequency range (5-20\,\hertz) and a high-frequency range (100-120\,\hertz), in equivalent temperature units for CMB fluctuations.
Uncertainties show the dispersion among 15-minute TODs.
The noise increase at higher frequencies is driven by intrinsic properties of the bolometers, as described in \citet{marriage:2006}, \citet{zhao/etal:2008} and \citet{niemack/etal:2008}.
However, our signal band is limited to below 30\,\hertz~by the telescope scan speed and the beam size (see Figure \ref{signal_ps_2008}). Thus, the mid-frequency range is the best estimate of the instrument sensitivity.

As given in Table \ref{noise_summary}, the typical noise variance per detector is $0.62\pm0.04\,\milli\kelvin^2\,\second$.
The total noise can primarily be separated into in-band detector noise, aliased detector and readout noise, and photon noise, all of which add in quadrature.

To reduce the effects of noise aliasing, the detectors are first sampled at 15.15\,\kilo\hertz, then a digital four-pole Butterworth low-pass filter with a cutoff frequency of 122\,\hertz~is applied, and finally the data are resampled at 398.72\,\hertz.
The detector noise bandwidth is limited  to 8\,\kilo\hertz~by a 700\,\nano\henry~inductor in series with each TES.

To determine the effects of optical loading and aliasing, we performed a dark test in which we opened the cryostat, put a 4\,\kelvin~(reflective) cover over the detectors, and collected data at a variety of sampling rates.
The typical noise level in the dark was $0.46\pm0.14\,\milli\kelvin^2\,\second$, where the uncertainty shows the dispersion among detectors.
The units are equivalent to the ones in Table \ref{sensitivities}.
After fitting the noise as a function of sampling frequency, the in-band detector noise yielded $0.30\pm0.10\,\milli\kelvin^2\,\second$ and the total aliased noise contribution was $0.17\pm0.08\,\milli\kelvin^2\,\second$.
The latter includes aliasing from both detector and readout noise. 

The readout noise is dominated by SQUID noise and preamplifier noise. 
Fully sampled, the readout noise is estimated to be around $1.2\times10^{-4}\,\milli\kelvin^2\,\second$ (based on 50\,\mega\hertz~measurements), which is expected to increase by roughly a factor 400 when sampled at 15\,$\kilo\hertz$, reaching $\approx\!0.05\,\milli\kelvin^2\,\second$.
The SQUID noise aliasing was significantly reduced in season 2010 by reducing the readout bandwidth.
Taking all this into consideration, we estimate that slightly more than half of the total aliased noise contribution is aliased detector noise, which is consistent with the detector noise aliasing analysis presented in \cite{niemack:2008}.
The other half is SQUID noise.


The dark tests also revealed that the optical loading contributes $0.15\,\milli\kelvin^2\,\second$ of photon noise.
By measuring the saturation power of the detectors at different atmospheric conditions throughout the season, and comparing them to the saturation power in the dark, we found that the optical loading is 2.24\,$\pico\watt$ when the PWV is 0\,$\milli\meter$.
This loading is dominated by spillover contributions, emission from optics and dry atmosphere emission. 
The spillover contribution was reduced by 0.36\,\pico\watt~in season 2010 by adding a baffling structure around the secondary mirror.
In the same way, we determined that water in the atmosphere contributes another 0.7\,$\pico\watt/\milli\meter$ of loading,  so in nominal conditions (PWV = 0.49\,\milli\meter) the total optical loading is 2.59\,\pico\watt.
This contributes to the noise as photon noise: for a fully incoherent detector coupling to both polarizations, the noise in units of power squared per unit frequency is given by

\begin{equation}
\mathrm{NEP}^2 \approx 2\,h\nu P + \frac{P^2}{\Delta\nu},
\end{equation}

\noindent where $h$ is Planck's constant, $\nu = 149.2\pm3.5$\,\giga\hertz~is the central frequency\footnote{The central frequency for a Rayleigh-Jeans source.}, $\Delta\nu = 18.4$\,\giga\hertz~is the bandwidth, and $P = 2.59$\,\pico\watt~is the power absorbed by the detector \citep{zmuidzinas:2003}.
The first term in the equation corresponds to photon shot noise while the second term corresponds to noise from photons arriving in bunches, as is expected for thermal radiation when the occupation number is high.
This leads to a photon NEP of $3.0\times 10^{-17}\,\watt/\sqrt{\hertz}$, or roughly 0.17$\,\milli\kelvin^2\,\second$ in CMB temperature units, in agreement with our measurements.

\tabletypesize{\footnotesize} 
\begin{deluxetable}{rc}
\tablecolumns{2}
\scriptsize
\tablecaption{Noise contribution summary.}
\tablewidth{230pt}
\tablehead{\colhead{Noise source} & 
           \colhead{NET$^2$ [$\milli\kelvin^2\,\second$]}}
\startdata
In-band detector noise &  $0.30\pm0.10$\\
Aliased noise & $0.17\pm0.08$ \\
Optical loading & $0.15$  \\[1.5mm]
\hline
\rule{0cm}{3.5mm}
Total (typical) noise & $0.62\pm0.04$ \\
\enddata
\label{noise_summary}
\end{deluxetable}%

Table \ref{noise_summary} shows a summary of the different uncorrelated noise contributions that determine the system sensitivity.

\begin{figure}[tbp]
\begin{center}
  \includegraphics[trim = 30 0 15 0, clip, width = 85mm]{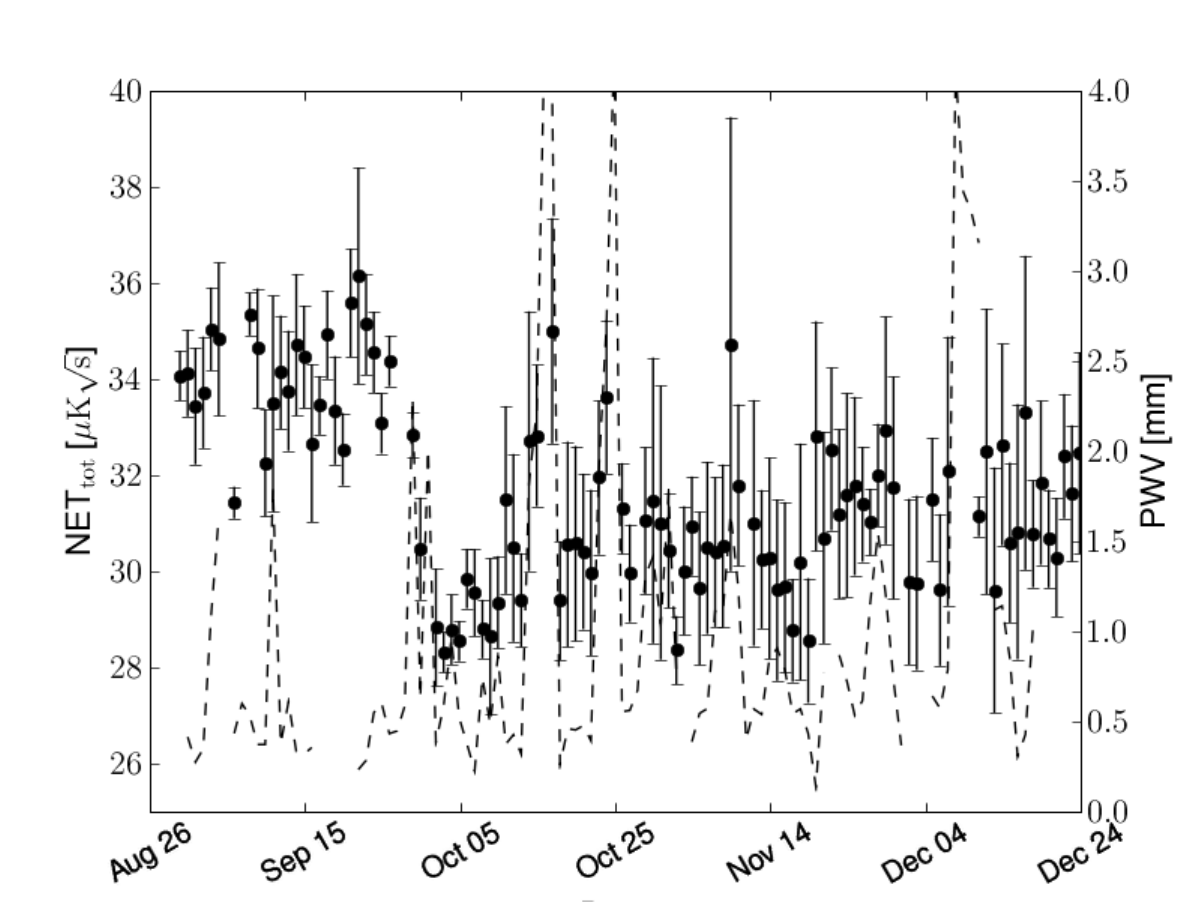}
\caption{\small{Total array NET during the 2008 season in the mid-frequency range (5-20\,\hertz) for \arone. The values are grouped in days and the error bars are the standard deviations within each day. 
  The PWV is also plotted for reference (dashed line).
  Values are calibrated to CMB equivalent units. 
  Before computing the sensitivity, eight multi-common modes, four row-correlated modes, four column-correlated modes, and the residual twelve modes with highest singular values were removed (see Appendix \ref{mode_removal} for details).
  The noise improvement after September 24 came from turning off some oscillating detectors, which were contaminating neighbors.} }
\label{season_nep}
\end{center}
\end{figure}

Figure~\ref{season_nep} shows the mean noise between 5 and $20\,\hertz$ for all data files in the 2008 season.
This is the total noise which goes into the maps and can be reduced only by increasing the observation time. 
Noise estimates from the maps can be found in \cite{marriage/etal:2011a} and \cite{das/etal:2011}.

\section{Systematic Noise}
\label{systematic_noise}

In addition to the atmospheric and random noises, the celestial signal is contaminated by several systematic noise sources. 
The important ones are thermal drift of the cryostat, mechanical accelerations which couple both optically and thermally (by causing detector temperature oscillations), electromagnetic pickup and magnetic pickup.  
All of these effects cause zero-lag correlations among different detector TODs, or can be directly measured using the dark detectors.
Detectors are distributed in 32 ``columns'' and 33 ``rows,'' where the last row contains the dark detectors.
Each column shares the same time-multipexed SQUID readout circuit \citep{deKorte/etal:2003} within which each row is read simultaneously.
Thus, there is one dark detector per readout circuit, serving to assess systematic noise from it.
Moreover, the high redundancy of live detectors across the array can also be used to assess systematic effects (see Appendix \ref{mode_removal}).

\subsection{Thermal Drift}

The current signals from the bolometers are amplified in 100-SQUID series arrays (SA) operated at 3\,\kelvin~\citep{swetz/etal:2011}.  
Slow temperature changes in the SAs produce slow drifts in the TODs.
During the first 10 hours of the night, the SA temperature drifts down by $250\,\milli\kelvin$, and rises up by $150\,\milli\kelvin$ in the last 2 hours. 
In terms of equivalent sky temperature at \arone, this corresponds to a drift of nearly 4\,\kelvin in signal.
In frequency space, the drift imprints a $1/f$ signature on the data, which meets the detector noise level at a knee of nearly $1\,\hertz$. 
Despite small differences in the responses of different SA amplifiers, most of this signal appears as a common mode to all the detectors.
As the coupling occurs at the readout circuit, this signal is also present in the dark detectors, simplifying its identification and eventual removal.  

In most cases this signal is subdominant to the atmosphere signal, but when the PWV is low enough the thermal drift becomes comparably significant.
This is not evident in Figure \ref{stare_atm_ps} since the examples of low-atmospheric power get averaged down when grouping many TODs.

In contrast, drifts in detector temperature cause no measurable effect because it is servo-controlled to within less than $1\,\milli\kelvin$ error.

\subsection{Mechanical Accelerations}
\label{mechanical_accelerations}

\begin{figure*}
\begin{center}
  \subfloat[Before removing correlated modes.]{\label{vibr_ww_before} 
  \includegraphics[trim = 65 50 160 110, clip, width = 82.8mm]{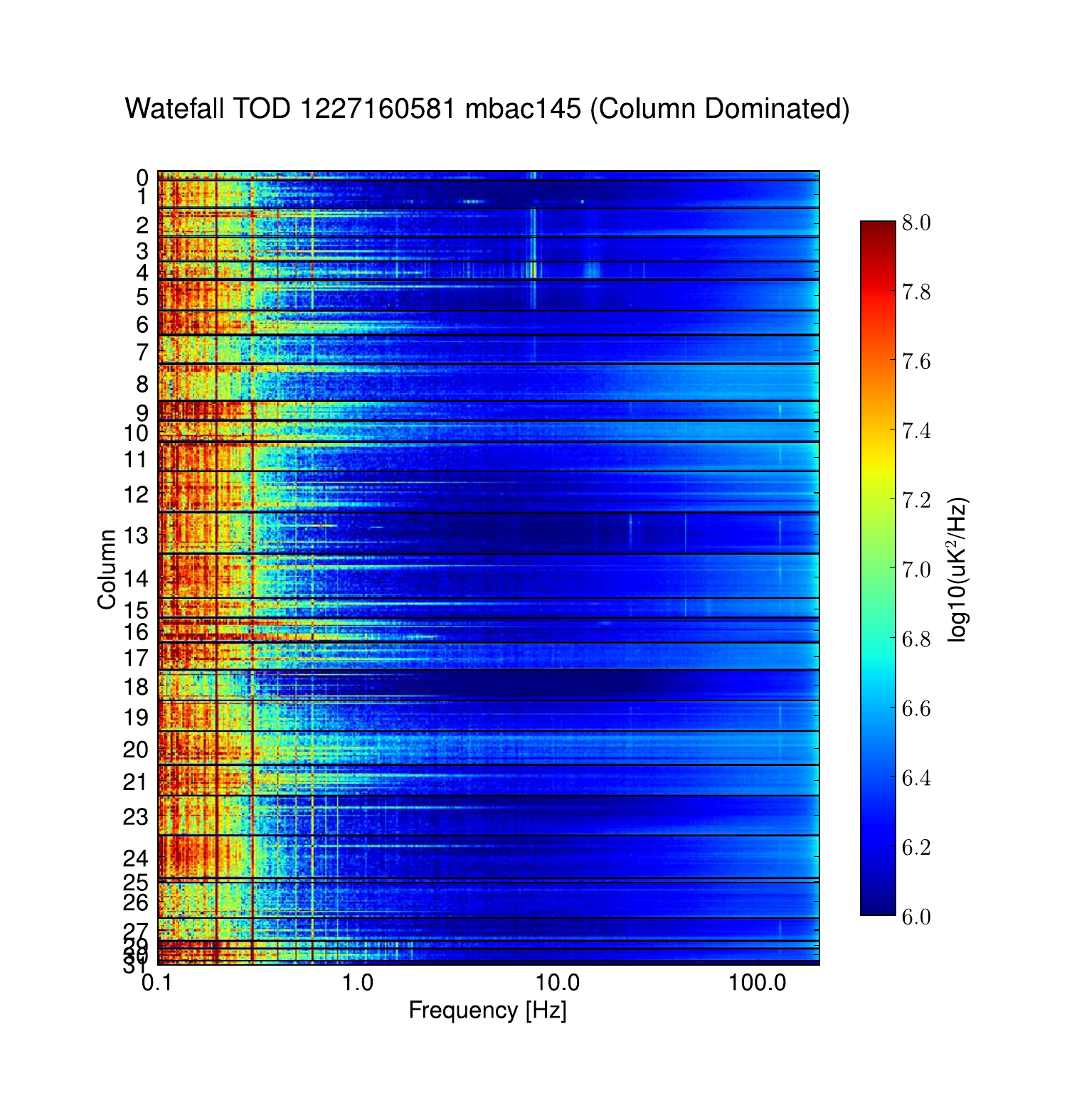}}
  \subfloat[After removing correlated modes.]{\label{vibr_ww_before} 
  \includegraphics[trim = 98 50 60 110, clip, width = 94mm]{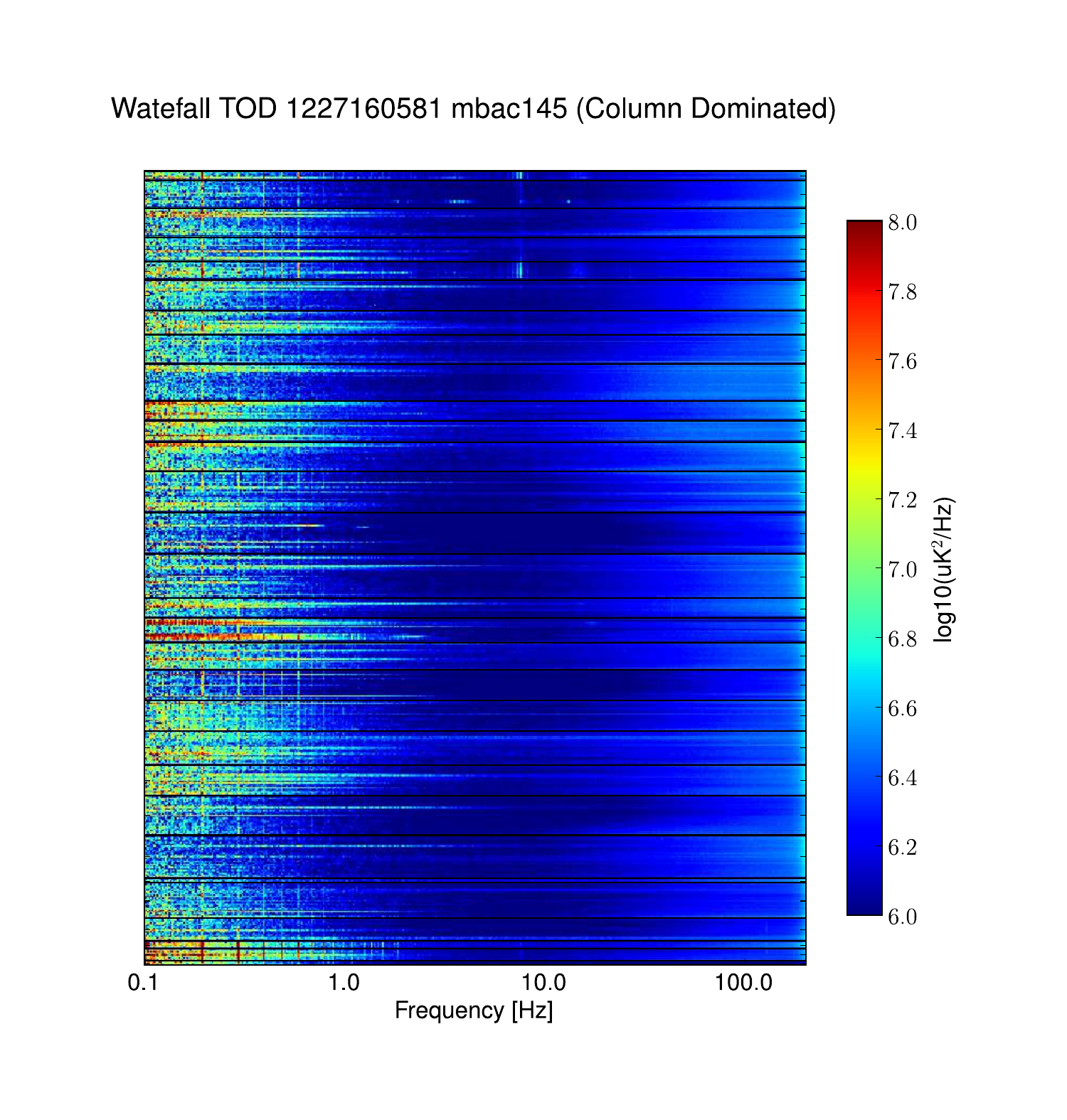}}
\caption{\small{Frequency-space waterfall plot containing the power spectrum from all live detectors from the \arone~band, before and after removing 20 correlated modes.
This TOD was obtained under nominal scanning conditions ($f_{scan} = 0.1\,\hertz$), on November 20, 2008.
The data are calibrated to CMB temperature equivalent units.
Each horizontal line in the 2D plot represents the power spectrum of a single detector.
The black horizontal lines separate detectors from different columns.
Most of the low frequency lines are scan harmonics.
They are explained by a combination of effects including spatial variations in the atmosphere brightness through the nearly triangular scan pattern, thermal oscillations of the cryostat and coupling layer oscillations.
The thick lines around 5\,\hertz~(see columns 2--5) are also present in stare observations.
They are thought to be resonant frequencies of the system.
Note that the science band is between about 1 and 20\,\hertz.
After removing 20 correlated modes from every detector TOD, harmonic features are significantly reduced, as is the low frequency power from the atmosphere and from thermal drifts.
When making maps, the correlated modes are identified and properly de-weighted to produce an unbiased map solution.
}}
\label{vibr_ww}
\end{center}
\end{figure*}

Mechanical accelerations, occurring mainly at the scan turnaround, can couple optically and thermally to the detector and result in an undesired instrumental response. 
The optical coupling can be produced by a mechanical motion of the detector coupling layer, which is a $40\,\mu\meter$ thick silicon layer placed $100\,\mu\meter$ away from the detectors for optical impedance matching.
The coupling efficiency is sensitive to the distance between the coupling layer and the detectors, so small vibrations can cause significant effects, especially at higher frequency bands where the coupling layer is thinner and the coupling efficiency is more sensitive to changes in the distance.  
The effect is also larger near the center of the detector array, presumably because the coupling layer vibrates in its fundamental mode.
In the TOD, vibrations manifest themselves as a series of spikes visible in the common mode at the scan turnarounds, with opposite signs for opposite directions, leading to lines in the power spectra at several harmonics of the scan frequency.
In the \arone~``waterfall plot'' of Figure \ref{vibr_ww}, the vibrations contribute to the scan harmonics and some resonant lines at higher frequencies.
The latter are also seen for stare observations, and their central frequencies are shared among all three arrays, suggesting that they correspond to natural frequencies of the whole system. 
To mitigate the mechanical effect, the turnaround acceleration was reduced from its value in the 2007 season to the value shown in Table~\ref{scan_parameters} after October 8, 2008.
The coupling layers for \artwo~and \arthree~were removed after the 2008 season.

Thermal perturbations of the cryostat, revealed in spectral analysis of the bath temperature, also add to the low frequency scan harmonics seen in Figure \ref{vibr_ww}.
This effect is column-dependent and differs from the coupling layer effect in that the signal is not stronger at the center of the array.

In general, high frequency spectral lines in the TOD from different detectors destructively interfere when projected into map space, mostly canceling out when many observations are combined. 
On the other hand, low frequency harmonics of the scan may produce non-negligible bar-like features in the maps perpendicular to the scan direction, if not treated properly.

\subsection{Electromagnetic Pickup}

Electromagnetic pickup couples to the readout circuit, producing various signatures in the data.
These signals appear correlated among subgroups of detector TODs, particularly among detectors from the same column or row.
We call these ``column'' or ``row'' correlations respectively, each corresponding to a different source of electromagnetic pickup.  
Narrow-band signals appear correlated among detectors within the same column.
They couple in somewhere after the first SQUID stage of the readout circuit, as the subsequent circuitry is shared by the detectors in a column.
On the other hand, broad-band signals appear correlated among detectors within the same row, or even a few rows apart.
We believe these are caused by rapidly varying signals, such as spikes from strong current transients: given our time-multiplexed readout scheme, in which the same row is read from all columns simultaneously, these signals become correlated among detectors in the same row.

\begin{figure*}
\begin{center}
  \subfloat[Column dominant. Magnetic pickup and atmosphere removed.]{\label{Mcm_sync_corr_mat_col} 
  \includegraphics[width = 88mm, trim = 0 0 0 27, clip = True]{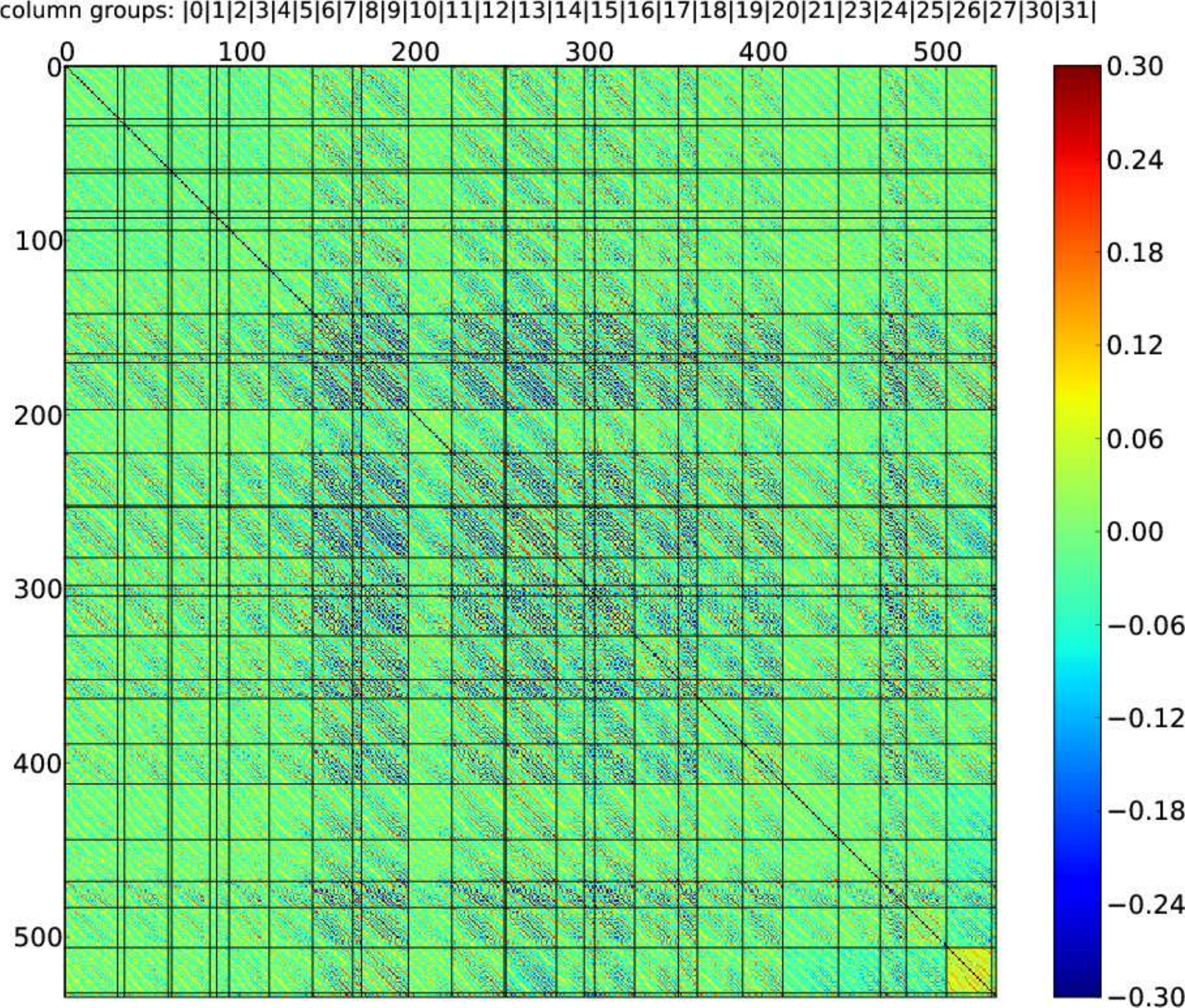}}
  \subfloat[Row dominant. Magnetic pickup and atmosphere removed.]{\label{Mcm_sync_corr_mat_row} 
  \includegraphics[width = 88mm, trim = 0 0 0 27, clip = True]{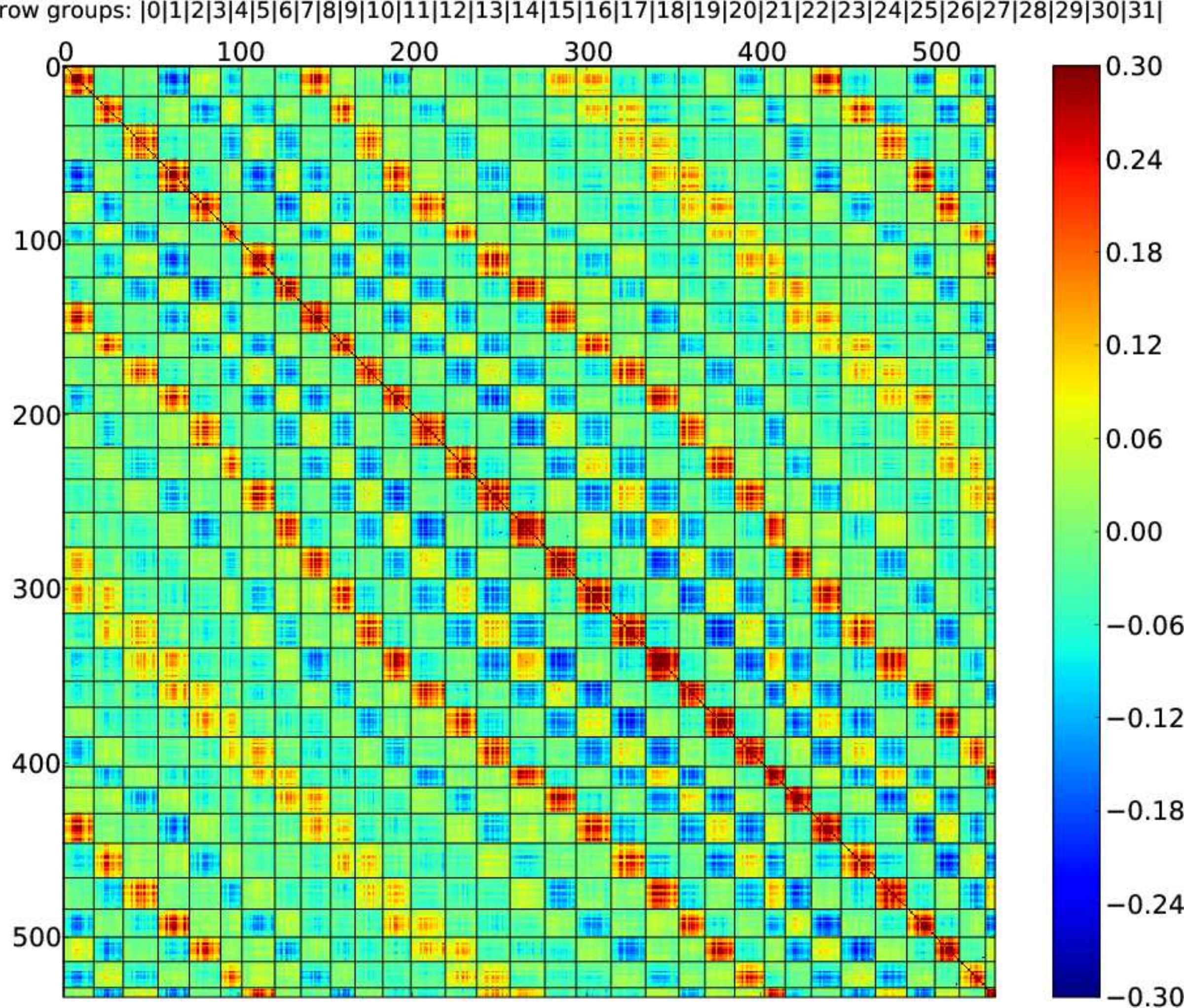}}
  \qquad
    \subfloat[Column dominant.  Magnetic pickup, atmosphere and dark modes removed.]{\label{Mcm_sync_dark_corr_mat_col} 
  \includegraphics[width = 88mm, trim = 0 0 0 27, clip = True]{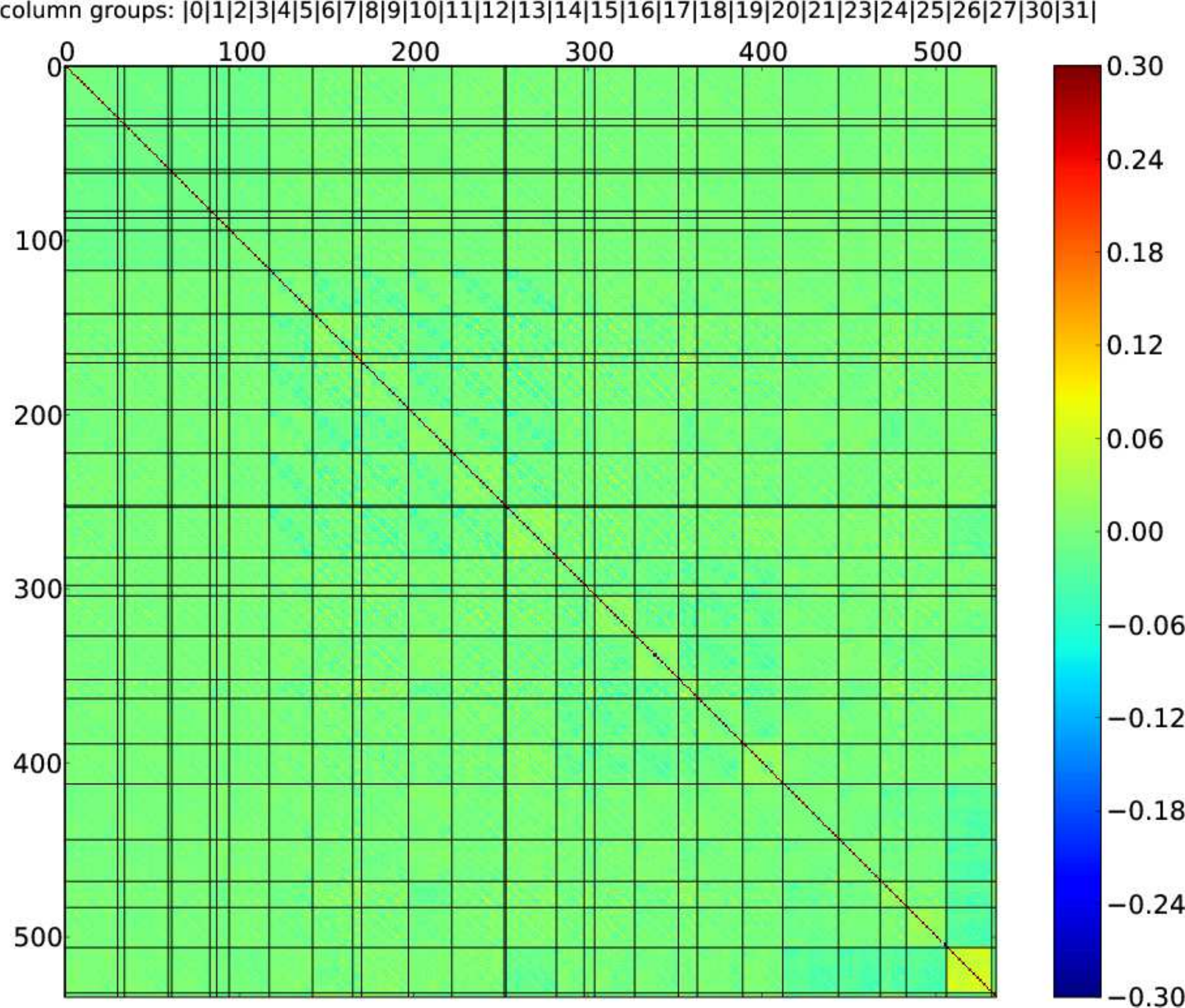}}
  \subfloat[Row dominant. Magnetic pickup, atmosphere and dark modes removed.]{\label{Mcm_sync_dark_corr_mat_row} 
  \includegraphics[width = 88mm, trim = 0 0 0 27, clip = True]{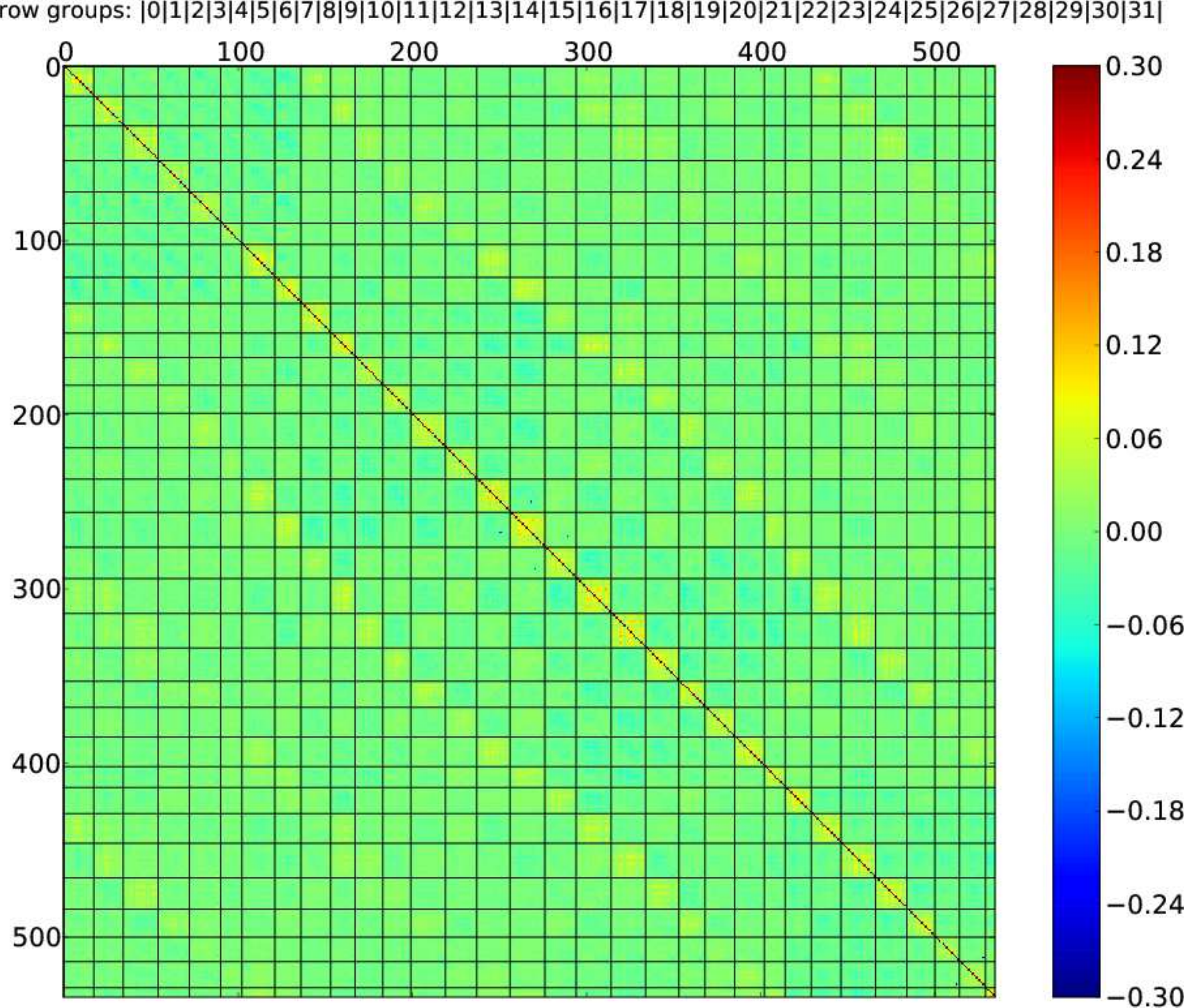}}
  \caption{\small{Correlation matrices for a 15-minute data stream obtained December 7, 2007. 
  The 551 active detectors are each numbered, starting from the upper left of the correlation matrix, in order of columns (left panels) or rows (right panels).
The remaining 473 detectors were cut from this particular data stream due to inadequate performance (see Sec. \ref{detector_cuts}).
The solid lines divide the 32 individual columns (left panels) or rows (right panels); the different widths between the solid lines reflect the varying number of active detectors in each column or row (cf.\ Fig. \ref{array_cut_distribution}). 
The magnetic pickup and the atmospheric signal have been removed from the time stream by subtracting a best-fit sinusoid for the former and by removing eight multi-common modes for the latter (see Appendix \ref{mode_removal}).
The top panels show these filters only, with a quality factor $Q=0.081$ (Eq. \ref{quality}). 
The bottom panels additionally have dark correlated modes removed, with an obvious large suppression of detector correlation and a quality factor of $Q=0.015$. }
  }
\label{sync_Mcm_corr}
\end{center}
\end{figure*}

Figure \ref{sync_Mcm_corr} shows the detector correlation matrix for a given 15-minute TOD file, where every element in the matrix corresponds to the correlation between the TODs of detectors $i$ and $j$, which index the elements of each axis. 
For instance, the diagonal elements correspond to $i = j$, so they are all equal to one.
In Figure \ref{Mcm_sync_corr_mat_col}, adjacent detectors on each axis belong to the same column in the array, with a black line separating detectors from different columns, while Figure \ref{Mcm_sync_corr_mat_row} is organized by rows.
We can clearly see the kinds of correlations described above affecting either columns or rows of detectors.  
As would be expected from a stochastic distribution of transients, the patterns seen in the row-correlated matrix change in time.
To quantify the correlations, a quality factor is defined as the mean of the squared off-diagonal elements of the correlation matrix:
\begin{equation}
Q = \frac{2}{N(N-1)}\sum_{i> j}^N{c^2_{i,j}},
\label{quality}
\end{equation}
where $c_{i,j}$ is the correlation between the TODs of detectors $i$ and $j$, and $N$ is the number of detector TODs.
Lower $Q$ factors indicate less correlated noise.

The column-correlated signals were significantly reduced by carefully shielding the system from electromagnetic noise in the environment. 
The row-correlated signals, instead, were found to be related to the switching power supplies feeding the readout electronics.
They were replaced by linear power supplies at the end of the 2008 season.

If not treated properly, broadband row-correlated signals can produce features in the maps, mostly seen as features perpendicular to the scan direction for that particular observation.

\subsection{Magnetic Pickup}
\label{magnetic_pickup}

The readout circuit of the detectors uses SQUIDs, which are very sensitive to magnetic fields.
Despite significant magnetic shielding, some residual pickup remains as the telescope sweeps through the Earth's magnetic field.
Although the scan is almost triangular, the observed signal is chiefly sinusoidal, which can be explained by eddy current losses and hysteresis in the magnetic shielding.
The magnetic signal amplitude and phase are fairly stable across detectors in the same column, but differ significantly between columns.
We believe these phase shifts are related to the complexity of SQUIDs and electrical loops in the system.
A typical amplitude of this signal is equivalent to nearly $T \simeq7\,\milli\kelvin$ in CMB units.

As this is a readout-related signal, it can be measured using the dark detectors and removed by fitting and subtracting a sinusoid.
Moreover, the scan frequency is significantly below the science band and the magnetic pickup is heavily down-weighted in the mapping.

\section{Calibration}
\label{calibration}

Our calibration takes into account the detector properties and electronics; camera and telescope optics; and atmospheric conditions.
It consists of three steps.
The first adjusts for variations in detector responsivity due to changing atmospheric loading and atmospheric opacity.
Next, a detector-detector gain term is applied to account for relative variations in optical coupling across the camera.  
Finally, a single normalization is applied to all the data to account for the overall system efficiency for celestial sources. 
This procedure is similar to the one presented in \citet{switzer:2008}.

\subsection{Calibration Variations with Time}
\label{subsec:calTime}

The nightly variations in atmospheric loading and daily cryogenic cycling necessitate a rebiasing of the detectors at the beginning of each night's observations, which affects the detector response.
Additionally, the opacity of the atmosphere varies from night to night.
Changes in loading and opacity throughout the night produce smaller variations in the system response.

In order to account for the variations in system response between nights, a responsivity calibration is derived from the analysis of $I-V$ curves taken during the nightly rebiasing.
An $I-V$ curve is the relation between the output current response ($I$) of the detector's SQUID readout circuitry to a slowly ramping detector bias voltage ($V$).
For the \arone~band, the median deviation in responsivity between nights as measured by the $I-V$ curves is 3.0\%.

A second test, known as a bias step \citep{niemack:2008}, is performed three times per night to detect changes in detector responsivity through the night.
This is achieved by recording the detector response to a series of small, square-wave pulses applied to the detector bias voltage.
The responsivities can be obtained by analyzing the detector responses to this signal, and are found to be in agreement with responsivities obtained from the $I-V$ curve analysis.
For the \arone~band, the median deviation in responsivity over a night, as measured by the bias step, is 1.0\%.
For more information about the ACT detectors, the biasing routine and responsivity, see \citet{fisher:2009}, \citet{battistelli/etal:2008}, \citet{swetz/etal:2011} and \citet{zhao/etal:2008}.
The time dependence on the atmospheric conditions will be discussed below together with the overall system calibration.  

\subsection{Time Independent Relative Detector Calibration}
\label{subsec:calRel}

To determine the detectors' relative gain coefficients, we use the large common-mode signal provided by the variations in atmospheric brightness. 
This is analogous to a flat-field calibration done with an optical CCD.
For each 15-minute data file, we compute the gain factor that best fits the detector drift to the common mode drift.
The dispersion of this factor per detector over the season, averaged over all detectors, is better than 2\% rms for the \arone~array.
This includes the variability of the time-dependent calibration step.
We correct for this by multiplying each detector time stream by its corresponding gain factor.

\subsection{Overall System Calibration}
\label{subsec:calTot}

An overall system calibration is established by measurements of Uranus.  
By comparing the peak response of the instrument to the known brightness temperature of Uranus (taking into account the ratio of Uranus' solid angle, $\Omega_{\mathrm{U}} = 0.235\pm 0.010\,\nano\steradian$, to the telescope's solid angle at \arone, $\Omega_{\mathrm{ACT}} = 215.8\pm1.5\,\nano\steradian$), an overall calibration of the system response is obtained.
The statistical error derived from 30 measurements in the \arone~band is 1\%, while solid angle uncertainties are below 1\%.
The equivalent Rayleigh-Jeans temperature of Uranus is obtained by reprocessing the data from \cite{griffin/orton:1993}, in combination with WMAP-7 measurements of the brightness of Mars and Uranus \citep{weiland/etal:2011}.
It yields $103.5\pm6\,\kelvin$ at \arone, producing a 6\% uncertainty in the overall calibration.


The conversion to sky temperature also depends on the atmospheric transmission.
This is estimated with the ATM model \citep{pardo/etal:2001}, which uses PWV measurements from APEX corrected for the ACT site and other Atacama-specific parameters. 
When PWV measurements are not available, the season-average transmission ($\mathcal{T} = 0.976$) is used.  
For the \arone~band, the rms of the transmission during the 2008 season was 2.3\%. 
Given that large atmosphere loading can excite non-linearities in the detector responsivities, an extra degree of freedom is given to the planet fit as a function of PWV, producing a final correction to the calibration factor.

The overall map calibration is compared to the WMAP-7 year map, by correlating multipoles in the range $400 < \ell < 1000$, providing an uncertainty of 2\% in temperature \citep{hajian/etal:2011}.
This analysis shows that the Uranus calibration is a factor of 6\% lower than the WMAP calibration, which is within the expected uncertainty.

Putting all pieces together, the overall system calibration for \arone~in the 2008 season is
\begin{equation}
F(\mathrm{w}) = C\,\exp\left[ \left(\tau_d + \tau_w\,w + \tau_x\left(w - \overline{w}\right)\right)/\sin\theta_{\mathrm{alt}}\right],
\end{equation}
\noindent where $C = 19.41\,\kelvin/\pico\watt$ is the overall calibration factor, $\theta_{\mathrm{alt}}$ is the observation altitude, $\tau_d = 0.0093$ is the ``dry opacity,'' $\tau_w = 0.0190\,\milli\meter^{-1}$ is the ``wet opacity,'' $\tau_x = 0.0138\,\milli\meter^{-1}$ is the the extra degree of freedom, $w$ is the PWV measurement from APEX, corrected for the ACT site in millimeters and $\bar{w} = 0.44\,\milli\meter$ is a pivot PWV used in the fit. 
Here $\tau_d$ and $\tau_w$ are fixed parameters obtained from the ATM model, while $C$ and $\tau_x$ are free parameters.
If $\tau_w$ is allowed to vary, we recover the same value that is predicted by the ATM model \citep{hincks:2009}.





\section{Pointing}
\label{pointing}

The pointing solution is decomposed into relative pointing between detectors and absolute boresight pointing, both of which we discuss here. 

\subsection{Relative Detector Pointing}

The relative pointing between detectors was determined by modeling the beam in the time streams of planet observations.
This analysis made use of approximately 30 observations of Saturn at the normal CMB observing altitude of 50\fdeg5, in two azimuth ranges corresponding to the rising and setting of the planet.

The telescope scans are rapid enough that each detector samples the vicinity of the peak response to a planet several times in a single observation.  
It is thus possible to model the beam position and shape in two spatial dimensions, project this to a detector signal using the telescope encoder information, and fit the data in the time domain.  
For simplicity, we use a 2-dimensional Gaussian as the beam model.
The fit produces azimuth and altitude offsets for each detector relative to the telescope pointing encoders, as well as measures of the peak response, beam FWHM, and optical time constants.

When combining different planet observations, we first aligned them by applying an offset correction to each one.
The position of each detector is taken as the mean of the positions from all observations, after rejecting outliers.
This produces relative detector positions for each array that are in good agreement with design expectations \citep{fowler/etal:2007}.  
A comparison between the offsets for rising and setting observations (which differ in azimuth by approximately 95\degree) shows no significant rotation or shear of the array relative to the local altitude and azimuth axes.
This constrains the tilt in the telescope azimuth axis to be smaller than 1\arcminute~and indicates that rotation of the array with respect to azimuth is negligible.

The uncertainty in the relative pointing is no greater than 1\farcs5 for all three arrays. 
Given the large number of detectors in each array, this small uncertainty in the relative detector positions amounts to a  negligible contribution to the pointing error in the final maps.

\subsection{Absolute Boresight Pointing}

The azimuth and altitude of the telescope encoder readings must be corrected to account for their offset from the true boresight for each frequency band. 
The correction is different for each of the  four central CMB observation azimuths, namely the equatorial and southern stripes at rising and setting orientations.

The correction is obtained by projecting the data from a particular band and orientation into a map (after having applied the relative pointing solution described above), and comparing the positions of the bright point sources to catalogue positions.
The resulting offsets in equatorial coordinates are converted to offsets in the boresight position, and this procedure is iterated to ensure convergence.
The boresight offset varies by about 1\arcminute over the season, primarily in elevation.

The dominant source of error in the offset correction comes from estimating the centroid positions of the point sources in the maps, before matching them to the catalog positions.
This error scales inversely with the square root of the number of point sources available, which was 20 for the \arone~band, and with their signal to noise ratio.  
The uncertainty was found to be 2\farcs6 for the southern stripe and 5\farcs3 for the equatorial stripe at \arone.
These values were obtained by adding in quadrature the error in the fit from the rising and setting maps.

The telescope pointing is expected to vary slightly due to thermal deformation of the mirror structure.
This variation is estimated from observations of Saturn taken at nearly the same azimuth and altitude on different nights.
The rms pointing variation of such observations is 4\farcs3.
Since any pixel in the final season maps contains contributions from many different nights, this random variation does not contribute significantly to the pointing uncertainty.

Pointing deviations are significantly higher at dawn when the telescope temperature begins to change more rapidly, showing a trend that repeats every morning.
The drift begins nearly 1 hour after sunrise, reaching nearly 50\arcsecond~an hour later.
Data taken more than 1 hour after sunrise are not included in the maps.

On top of this, the altitude was observed to drift by about 20\arcsecond~over the course of the 120 day season, producing pointing errors that were correlated with right ascension. 
To remove this trend, linear corrections of 0\farcs2/day and 0\farcs15/day were applied to the rising and setting fields respectively.
These corrections have little effect on the maps other than to remove the small, RA-dependent pointing offset.

The net pointing uncertainty is thus dominated by the systematic uncertainty in the alignment of the rising and setting maps, along with some residual variation due to temperature-dependent mirror deformation.
Comparing the positions of bright point sources in the maps to their catalog positions, we estimate the pointing error in the final maps to be 4\farcs8, with no preferred orientation \citep{marriage/etal:2011a}.
Note that this uncertainty is much smaller than the beam size, and would cause errors of less than 1\% in the measured flux for point sources.


\section{Detector Time Constants}
\label{time-constants}

A detector's time response is limited by its electro-thermal properties.
We model their optical step response as an exponential decay with a time constant $\tau$.
A finite response time results in a small shift in the spatial position of a point source.
The shift depends on the scan direction.
In the fits for the relative pointing solution described above, time constants were included in the beam model as a single-pole low-pass filter in the time domain.
The time constants are measured with a precision of $\delta \tau \lesssim 0.5\,\milli\second$, which is an upper bound to their dispersion from the analysis of 30 Saturn observations.
The median time constant of the \arone~array is $1.9\pm0.2\,\milli\second$ ($f_{3dB} = 83.8\pm8.8\,\hertz$), with only a handful of detectors showing responses slower than $\tau = 10\,\milli\second$ ($f_{3dB} \approx\!15\,\hertz$).
For comparison, Equation \ref{f_tod_vs_ell} implies that $\ell = 10^4$ corresponds to 27\,\hertz.

\section{Data Selection}
\label{data_selection}
Data selection can be divided into two types: data file selection (all detectors) and single-detector TOD cuts.
The former determines the number of observing hours, while the latter determines the number of ``effective detectors'' within each data file, defined as the sum of the fraction of the time that each detector passed the data selection:

\begin{equation}
  N_{\mathrm{eff}} = \sum_{i=1}^{1024}{\frac{T_{i,\mathrm{uncut}}}{T_{\mathrm{total}}}},
\end{equation}

\noindent where $T_{i,\mathrm{uncut}}$ is the available time for detector $i$ after data selection, and $T_{\mathrm{total}}$ is the total time before data selection. 
The number of array-wide effective detectors can then be defined as the average of $N_{\mathrm{eff}}$ over all the available data files in the season.

The remainder of this section presents the data selection methods and statistics for these two types of data cuts.

\subsection{Detector Classification}

The detectors are classified in three groups:
\begin{itemize}
\item{Live Detector Candidates: Those that have the potential to be used for map making;} 
\item{Dark Detector Candidates: Those that do not couple to the sky (for example a broken pixel) but can be used to diagnose systematics. This includes the subset of the 32 dark detectors which work properly and the subset of defective detectors with working readout circuit;} 
\item{Broken Detectors: Those defective detectors that cannot be used to probe systematic effects.
They include those former live and dark detectors with defective readout circuits, and slow detectors ($\tau < 10$\,ms). Many of them must be turned off while observing to prevent them from interfering with other detectors.}
\end{itemize}

Several methods were used to classify detectors into these categories.  These include assessing correlation with the atmospheric signal (by far the largest signal), searching for consistently oscillating detectors,\footnote{We have implemented automatic methods to identify oscillating detectors on the fly and disconnect them.} and finding biasing problems.
Table \ref{array_summary} gives a summary of the number of detectors in each of the three groups for each array.

\begin{deluxetable}{rc}
\tablecolumns{2}
\tablewidth{230pt}
\tablecaption{\footnotesize{Summary of number of detectors in each of the three groups, number of detectors cut by each criterion and effective number of detectors for \arone~in 2008. Errors are the standard deviation of the number of detectors.}}
\tablehead{ & \colhead{Number} }
\startdata
\sidehead{\em Detector classification}
	Live Candidates 	& 795  \\
	Dark Candidates 	& 128  \\
	Broken Detectors 	& 133  \\
	Total                  		& 1056  \\
\sidehead{\em Cuts by criterion applied to live detectors}
	Calibration  	&  $3  \pm 4$ \\
	Drift            	&  $95 \pm 61$\\
	Correlation  	& $8 \pm 16$ \\
	Gain 		& $2 \pm 8$ \\
	Mid-F. Noise 	& $41 \pm 33$ \\
	HF Rms 	         & $15 \pm 12$ \\
	HF Skewness   	& $3 \pm 4$ \\
	HF Kurtosis	& $7 \pm 9$ \\
	Scan 		& $2 \pm 9$ \\
	Glitch         	& $26 \pm 23$ \\
\sidehead{\em Effective detectors}
	& $593 \pm 95$ \\
\enddata
\label{array_summary}
\end{deluxetable}%

Once the live and dark detector candidates have been identified, a number of possible pathological behaviors may still justify removing part of the data from the reduction pipeline. 

\subsection{Data File Selection}

Out of the total number of data files acquired, we rejected files for the following reasons and in the following order:
\begin{itemize}
\item{Files that correspond to planet observations and other calibration or engineering tests;}
\item{Data taken more than one hour after sunrise, to avoid pointing and beam errors caused by telescope deformations as it was thermally settling;}
\item{Files with fewer than 400 effective detectors, as they were considered likely to be pathological;}
\item{Bad weather: PWV greater than $3.0\,\milli\meter$ (transmission below 90\%);}
\item{Poor cryogenic performance: Detector base temperature more than $7\,\milli\kelvin$ above the nominal temperature, or when it changed more than $1\,\milli\kelvin$ within the 15-minute file;}
\item{Poor calibration: If the relative gain dispersion of the detectors was more than 10\%;}
\item{Data for which the analysis software failed.}
\end{itemize}

The final amount of data available for analysis is summarized in Table \ref{available_data}.

\begin{deluxetable}{lrr}
\tablecolumns{3}
\tablewidth{230pt}
\tablecaption{\footnotesize{Data file selection in 2008 at \arone.}}
\tablehead{\colhead{Type} & \colhead{Obs. Hours} & \colhead{\%}}
\startdata
	Calendar                          			& 3264\,\hour 	& 100\% \\
	Total observation           			& 1423\,\hour 	& 43.6\% \\
	Total survey                     			& 1260\,\hour 	& 38.6\% \\
	Later than 1\,\hour~after sunrise  	& -198\,\hour 	& (6.0\%) \\
	Low effective detectors 			& -156\,\hour 	& (4.8\%) \\
	High PWV conditions     			& -41\,\hour 	& (1.2\%) \\
	Cryogenic problems  			& -28\,\hour 	& (0.9\%) \\
	High gain dispersion      			& -10\,\hour 	& (0.3\%) \\
	Other                                  			& -13\,\hour 	& (0.4\%) \\
	Uncut South                                           & 772\,\hour	& 23.7\% \\
	Uncut Equator					& 44\,\hour	& 1.3\% \\
	Uncut Total Survey           			& 816\,\hour 	& 25.0\% \\
\enddata
\label{available_data}
\end{deluxetable}%
\tabletypesize{\normalsize} 

\pagebreak
\subsection{Detector Cuts}
\label{detector_cuts}
Detectors are affected by sporadic pathologies. Depending on the kind of pathology, it may be necessary to reject a section or the full length of a detector TOD from a given data file.

The main causes for these pathologies are quantum jumps of the magnetic flux of a SQUID in the readout circuit (V-$\phi$ jumps); excessive detector noise; conducted noise from oscillating detectors; excessive electromagnetic pickup; and mechanical contamination, which can be optically or thermally coupled into the signal.

The following tests were performed over 15-minute data files to detect these pathologies:
\begin{itemize}
\item{{\it Drift test:}
This probes low-frequency deviations of a detector TOD from the atmospheric signal.
The data are first low-pass filtered above $50\,\milli\hertz$ and calibrated into units of power, as described in \S\ref{calibration}.
Then the thermal drift is removed by de-projecting both the dark detector common mode and the housekeeping temperature of the detectors.
Finally, the atmosphere signal is removed by de-projecting eight multi-common modes, as explained in Appendix \ref{mode_removal}. 
The drift error is defined as the standard deviation of the residual data.
Detectors in \arone~are cut as outliers if their drift error is higher than $0.35\,\femto\watt$.
}
\item{{\it Correlation test:}
The Drift test is complemented by finding the correlation between the detector drift and array common mode drift for TOD frequencies below $50\,\milli\hertz$. Detectors that correlate less than 98\% are excluded.
}
\item{{\it Gain test:}
The Drift test assesses the shapes of the TODs, but not their amplitudes.
The relative detector gain can be quantified by the factor that best fits the atmospheric drift to every single TOD at frequencies below $50\,\milli\hertz$.
For this the atmospheric drift is estimated as the array common mode after removing the thermal contamination.
Detectors are cut whenever their gains differ from the mean by more than 15\%. 
}
\item{{\it Mid-frequency noise test:}
Some pathologies, mainly associated with mechanical contamination, become prominent at frequencies between 0.3 and 1.0\,\hertz, below the ``science band.''
To isolate pathological detectors, we band-pass filter the data below and above those frequencies, de-project the array common mode (also filtered), and obtain the standard deviation of the residuals, which we call mid-frequency error.
Detectors are cut whenever their mid-frequency error is more than 2.5 times the median value for the data file.
}
\item{{\it High-frequency noise tests:}
At frequencies above 3\,\hertz~the data start being dominated by detector noise, which is chiefly Gaussian.
Non-Gaussianities above this frequency are mostly caused by electromagnetic pickup, conducted electrical noise or opto-mechanical perturbations.  
To isolate them, we high-pass filter the data below $5\,\hertz$ and then test for non-Gaussianities by computing the standard deviation, skewness and kurtosis.
This is done within sections of the TOD of one scan period, with a characteristic length of $10.2$\,\second.
Using the transformation proposed by \citet{d'agostino/Belanger:1990}, the last two statistics yield a normal distribution in the case of Gaussian noise, which is verified for most of our data.
Outliers are identified by how much they deviate from the mean. 
Only sections of the TOD with noise rms lower than $0.9\,\femto\watt$ are kept.  
}
\item{{\it Scan test:}
Motion defects and encoder errors are also detected and cut, affecting sub-sections of the TOD. 
This is done by examining the azimuth time stream and searching for interruptions in the scan pattern.
}
\item{{\it Glitch test:}
The data are also affected by spike-like glitches, for example from cosmic rays. Spikes larger than 10 times the noise rms are cut, including a $0.5\,\second$ buffer on either side. 
Also, whenever two such cuts are separated by less than $5\,\second$ they are stitched together into a single cut.
If more that 5 glitches are found in a single detector TOD, then the whole detector TOD is cut.
}
\item{{\it Calibration Bolometer test:}
During 2008 observations a calibration bolometer was used to load the detectors with radiation of roughly 400\,\milli\kelvin~every 24 minutes, each event lasting for 1.3 seconds.
For these, a window of nearly $3\,\second$ is excised around the event.\footnote{The calibration bolometer was not used after 2008.} We included this within the "Glitch" cuts in Table \ref{array_summary}.
}
\end{itemize}

As a general rule, if more than 20\% of the detector TOD would be cut, then the full detector TOD is cut instead.

Dark detectors were selected in an analogous but simplified way.
In this case, only full detector TOD cuts were performed. 
The cut criteria were the drift error of the dark detectors, the gain with respect to the dark common mode\footnote{Notice that the common mode of the dark detectors is driven by the thermal drift of the cryostat, which is the second largest signal in the live detector data.} and the noise rms, all given in raw data units.

\subsection{Data Selection Results}

After applying all the selection criteria given above, and considering only the data files available for analysis, the average number of effective detectors was $593 \pm 95$~in the 2008 season at \arone.
The error here is the standard deviation over different data
files.

\begin{figure}
\begin{center}
  \includegraphics[trim = 20 60 20 50, clip, width=85mm]{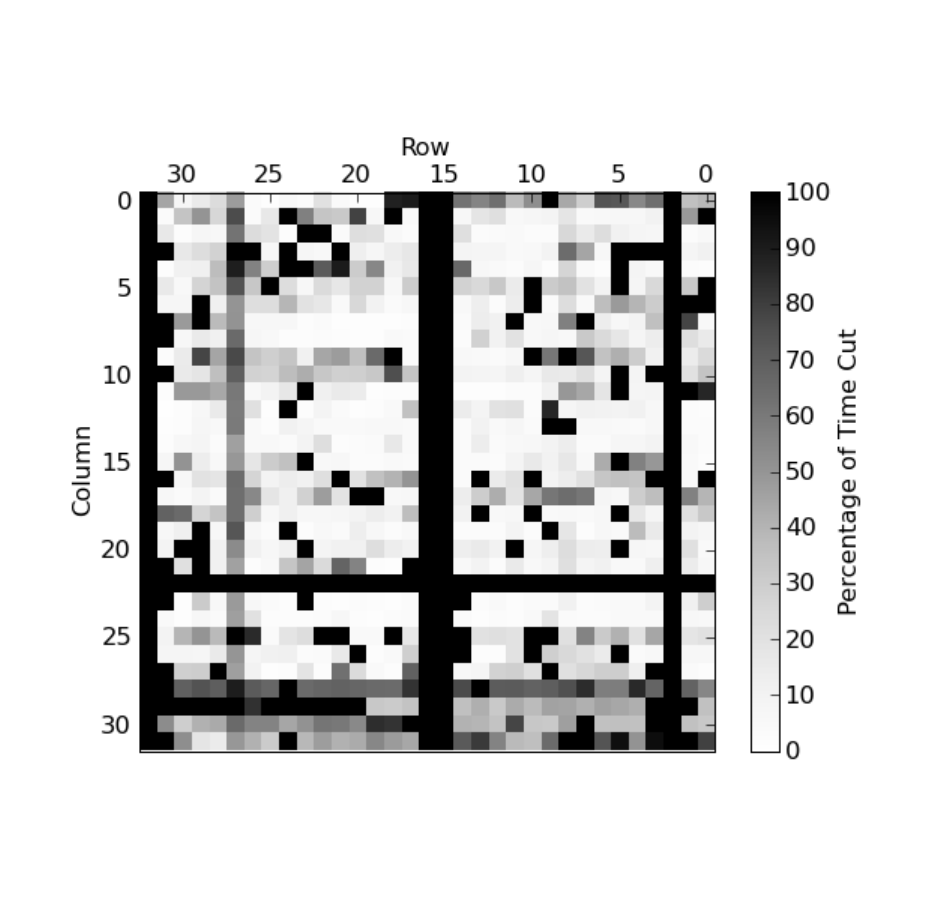}
\caption{{\small Percentage of time that detectors were cut across the \arone~array. 
Each small square represents a single detector. 
They are oriented as the array is projected on the sky. 
Notice that some rows and columns are always cut, which is mainly due to problems in the biasing (rows) and readout circuits (columns).}}
\label{array_cut_distribution}
\end{center}
\end{figure}

\begin{figure}
\begin{center}
  \includegraphics[width=85mm, trim = 20 0 10 0, clip]{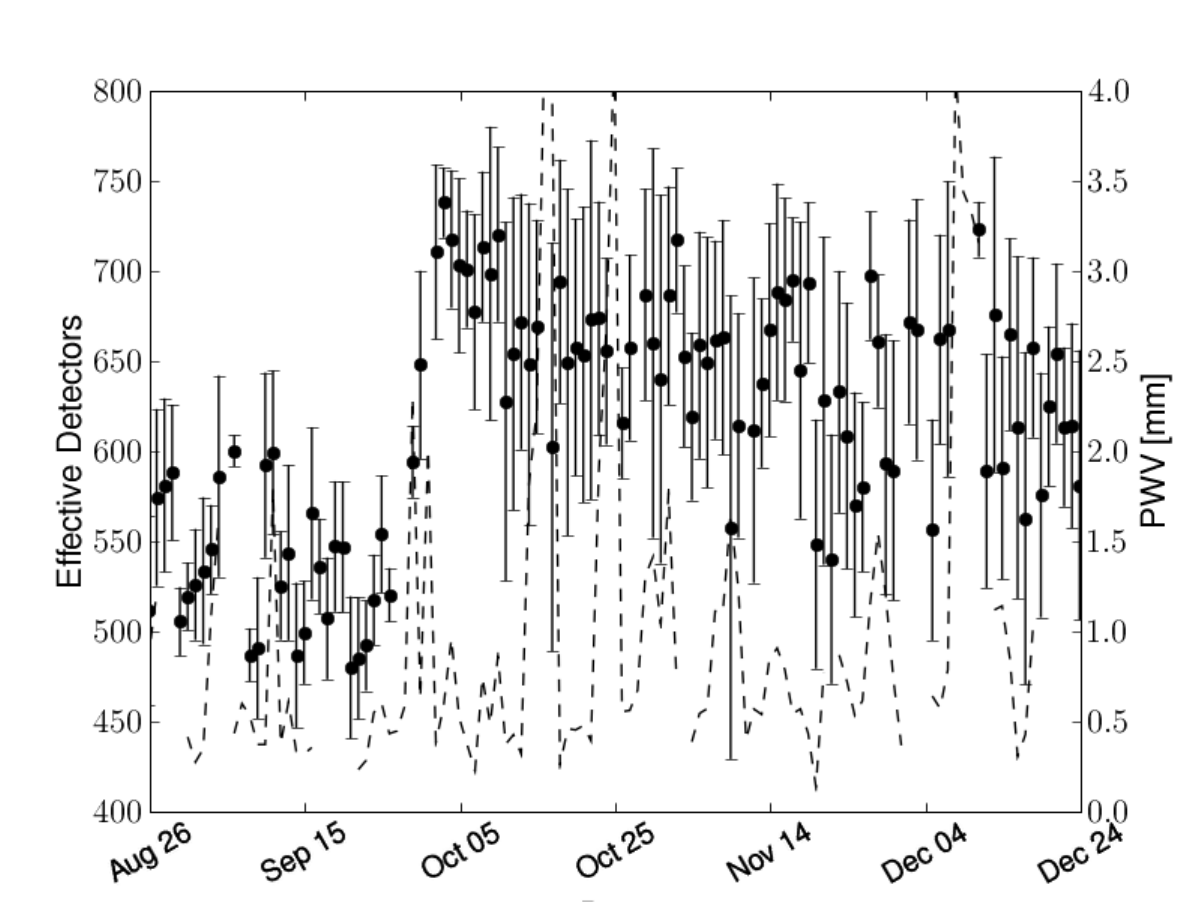}
\caption{\small{Effective number of detectors in the \arone~array in the 2008 season. 
Circles denote daily averages and the error bars are the standard deviation within that day. 
The PWV level is shown with the dashed line. 
The increase in the number of effective detectors starting in September 24 occurred after some oscillating detectors, which had been contaminating other detectors, were turned off.}}
\label{effDets_season}
\end{center}
\end{figure}

Table \ref{array_summary} shows a summary with the number of detectors in each of the three detector groups for the \arone~array, as well as the cut contribution from the eight main selection criteria: drift, gain, mid-frequency noise, noise rms, noise skewness, noise kurtosis, and partial cuts with more than 20\% of the TOD cut. 
The bottom line is the average number of effective detectors in the season.
Figure~\ref{array_cut_distribution} is a diagram of the \arone~array showing the fraction of the time that each detector is uncut. 
Figure \ref{effDets_season} shows the daily number of the effective detectors during the 2008 season, along with the weather conditions indicated by the PWV. 

On top of these cuts, a small fraction of the remaining data is cut during the mapmaking process, as described in the following section.

\clearpage
\section{Map making}
\label{map-making}

\subsection{Mapping Essentials}

To make maps from ACT data, we solve for the best-fit sky given the noise in the data.
In particular, we find the sky map that minimizes $\chi^2$ given a model for the noise, and a model for what the data should look like:
\begin{eqnarray}
  \mathbf{d} &=& \mat{M}\mathbf{x}+\mathbf{n}.
\end{eqnarray}

\noindent Here $\mathbf{x}$ is the model for which we wish to solve, $\mat{M}$ describes how the data depend on the model, and $\mathbf{n}$ is the particular realization of the noise in the ACT data.
Traditionally, $\mathbf{x}$ is a vector whose components are the sky map pixels and $\mat{M}$ is the pointing matrix.
In its simplest conceivable form, each data point sees a single pixel in the map, so each row of $\mat{M}$ (corresponding to a single data point) has a single 1 in the column corresponding to the map pixel at which it was pointed.
However, our model for the data, $\mathbf{x}$, can contain many contributions in addition to the sky map: components include atmospheric noise, correlated signals in the data, point source fluxes, and missing (cut) data samples as discussed below.
The mapping formalism easily generalizes to cover multiple components as long as the data depend on them linearly:
\begin{eqnarray}
  \mat{M} &=& \left[\begin{array}{cccc}\mat{M_1} & \mat{M_2} & \cdots & \mat{M_c} \end{array}\right], \\
  \mathbf{x} &=& \left[\begin{array}{c}\mathbf{x}_1 \\ \mathbf{x}_2 \\ \vdots \\ \mathbf{x}_c \end{array}\right].
\end{eqnarray}
With an estimate of the noise covariance $\mat{N} \equiv \left <\mathbf{n} \mathbf{n}^T \right > $, which is a $n_\mathrm{sample}$ by $n_\mathrm{sample}$ matrix ($n_\mathrm{sample} \simeq 10^9$ ), we wish to find the model that maximizes the likelihood function
\begin{equation}
\mathcal{L} =  \exp\left[-\frac{1}{2}(\mathbf{d} - \mat{M}\mathbf{x})^T \mat{N}^{-1}(\mathbf{d} - \mat{M}\mathbf{x})\right],
\end{equation}
\noindent where $\mathbf{d}$ is a vector containing all the data samples.
\noindent The standard linear least-squares solution is
\begin{equation}
  \mat{M}^T\mat{N}^{-1}\mat{M}\mathbf{x} = \mat{M}^T\mat{N}^{-1}\mathbf{d}.
\end{equation}

\noindent The matrix $\mat{M}^T\mat{N}^{-1}\mat{M}$ is too big to be practicably inverted directly (we typically have $10^7$ map pixels), so we instead iteratively solve the least-squares equation for $\mathbf{x}$ using a Preconditioned Conjugate Gradient (PCG) scheme \citep{wright/etal:1996, hinshaw/etal:2003, press/teukolsky/vetterling:NRC:3e}.
Preconditioning involves introducing $\mat{P}$, an approximate inverse of $\mat{M^T} \mat{N}^{-1} \mat{M}$, in order to speed up convergence of classic Conjugate Gradient, and solving the better-conditioned system:
\begin{equation}
  \mat{P}\,\mat{M}^T\mat{N}^{-1}\mat{M}x =  \mat{P}\,\mat{M}^T\mat{N}^{-1}\mathbf{d}.
  \label{eq:mapeq}
\end{equation}


To map ACT data, we use the mapmaking code Ninkasi and run on the Scinet General Purpose Cluster (GPC) \citep{Loken/etal:2010}.
The mapmaking algorithm for the maps in this release is improved from that used in previous papers \citep{fowler/etal:2010, marriage/etal:2011a, marriage/etal:2011b, dunkley/etal:2011, das/etal:2011, hajian/etal:2011, sehgal/etal:2011} in four ways:
1) rather than solving for detector-correlated noise (including atmospheric noise) explicitly, we put the correlations in the noise matrix;
2) we explicitly solve for cut data (time-stream gaps);
3) we subtract models for the point sources in the ACT maps directly from detector time streams;
and 4) we re-estimate the noise after creating an initial map to remove signal-induced bias in the noise estimation. 

The mapping is done in a cylindrical equal-area projection with a standard latitude of $\delta=-53\degree.5$ and pixels of $30\arcsec\times 30\arcsec$, roughly one-third of the beam FWHM\@.

\subsection{Pre-processing}

In addition to data selection, calibration and pointing, there are a few other pre-processing steps that are done to the data before solving for the maps.


We remove the median of each detector time stream for each 15-minute period, and subtract a single array-wide slope across the period so that the ends of the time streams roughly line up.
We do this to reduce ringing in Fourier transforms and to facilitate searching for correlations among the detectors.  
We linearly interpolate across gaps in the data, such as those arising from cosmic ray hits (again, to reduce Fourier artifacts: the data in the cuts are otherwise not used).
We deconvolve the time streams by the anti-alias filter (described in \S\ref{random_noise}) and the detector time constants (discussed in \S\ref{time-constants}).

Next, the non-optical contamination signals are reduced using the dark detectors.
This is done as follows: we take the time streams from each dark detector and subtract a mean and a slope.
We then high-pass filter the dark time streams, filtering out any signal below 5\,\hertz.
Most of the remaining signal is in only a few independent modes.
We take those corresponding to the seven largest eigenvalues of the resulting covariance matrix and do a linear least-squares fit to the live detector data, which had been similarly-processed (having removed the mean and slope, and high-pass filtered).
Finally, we subtract the reconstructed fit from the data.
We can do this because the modes subtracted are not correlated to the sky signal.

In \citet{fowler/etal:2010}, we downsampled the timestreams from 400 Hz to 200 Hz, using a time-domain triangular kernel of the form [0.25 0.50 0.25].  The ACT signal band (nearly $30\,\hertz$ according to Sec.~\ref{signals}) is well away from the downsampled frequency limit.  We find that downsampling does affect the raw power spectrum at up to several percent for $\ell > \sim 5000$.  This should be mostly corrected for when using a beam measured using maps derived from downsampled data.  However, we also find that the SNR on point sources is 2\% higher for non-downsampled maps.  Consequently, we do not downsample the timestreams in the results presented here.

Also in contrast to \citet{fowler/etal:2010}, we carry out one further step in the time domain:
With the 30\arcsecond~pixels somewhat undersampling the ACT 1\farcm37 beam, we find that, to get percent-level accuracy in source fluxes, we must deal with the bulk of the source flux directly in the time streams.
To do this, we make first-pass maps in which sources are found and their fluxes estimated.
Simulations show that these fluxes are typically accurate to a few percent, with source flux systematically underestimated by approximately 4\%.  
We then take these source fluxes and model them directly in the time streams using the full ACT beam, and subtract the model from the time streams.
We do a simple source-only projection into a map, and add this to the (mostly) source-free maximum likelihood map.
We find that, in simulations, this recovers mean source fluxes to 1\% accuracy, at which point it is subdominant to the calibration uncertainty.
We do no additional filtering in the time domain.

\clearpage
\subsection{Noise Modeling}

The noise structure is modeled in both frequency and time domains.
We include, as a term in the noise, a time domain windowing of the first and last 20 seconds of each time stream of the form $\left (1-\cos x \right)/2$, again to reduce Fourier ringing.
We then search for correlations across the data by examining their covariance matrices, split at 4 Hz ($\ell \simeq 1500$).
We find all eigenvectors in the low-frequency covariance matrix with corresponding eigenvalues greater than $3.5^2$ times the median (the eigenvalues of the data covariance matrix $\Delta^{T} \Delta$ are the squares of the corresponding data singular values, so this corresponds to an amplitude of 3.5 times the median in the time streams).
We then project those eigenvectors out of the high-frequency covariance matrix, and again find all eigenvectors with eigenvalues larger than $3.5^2$ times the median in the remaining high-frequency matrix.
We find that the maps are not particularly sensitive to the exact threshold values chosen.
This procedure typically finds 15 to 20 low-frequency eigenvectors and one or two high-frequency ones.
The eigenvectors are response patterns of correlations across the array, and usually correspond to things like a common mode, gradients across the array, and the row-correlated noise.
They provide the linear combination of detector time streams needed to produce a correlated mode, represented by the vector $\mathbf{\hat{v}}$ in Equation \ref{linear_comb}.
This projection corresponds to removing around 20 modes out of the approximately 600 available.

With the shapes of the array correlations in hand, we can use them to complete the description of the noise.
We solve for the correlated modes corresponding to the eigenvectors and subtract them from the time streams.
We then Fourier transform the (de-correlated) detector time streams and the correlated modes, and model them using frequency bins.
In each frequency bin, we find the average power in each detector time stream and in each correlated mode.
If the detector noises are denoted by the diagonal matrix $\mat{N}_{D}$, the detector covariance eigenvectors by the $n_\mathrm{detector}$ by $n_\mathrm{mode}$ matrix $\mat{V}$ and their corresponding bin-wise noise powers by $\mat{N}_{V}$, then the bin-wise Fourier-space noise is simply $\mat{N}_{f} \equiv \mat{N}_{D}+\mat{V}\,\mat{N}_{V}\,\mat{V}^T$.
Here $\mat{N}_f$ is an $n_\mathrm{detector}$ by $n_\mathrm{detector}$ matrix acting on a single frequency bin.
This matrix can be quickly inverted using the Sherman-Woodbury formula (\cite{duncan:1944}; see \cite{hager:1989} for a review).
If we denote the time-domain edge tapering operator by $\mat{W}$ and the Fourier transform operator by $\mat{F}$, then our entire noise inverse becomes 
\begin{equation}
\mat{N}^{-1} \equiv \mat{W}^{T}\,\mat{F}^{T}\,\mat{N}_{F}^{-1}\,\mat{F}\,\mat{W},
\end{equation}
where the operator $\mat{N}_{F}^{-1}$ is conformed by all the $\mat{N}_f^{-1}$ such that it acts on every frequency bin in the Fourier transform.
In the previous equation, every operator can be represented by an $n_\mathrm{sample}$ by $n_\mathrm{sample}$ matrix, in which detector operations like $\mat{N}_f$ are expanded such that all samples within a single detector are treated in the same way.
To minimize our sensitivity to any low-frequency non-Gaussian component of the atmospheric noise, we taper the Fourier weights at frequencies below 0.5\,\hertz, with the weight explicitly set to zero below 0.25\,\hertz. 
We note that the noise in the mapmaking equation (Eq. \ref{eq:mapeq}) can be interpreted as a set of weights.
The map is {\it optimal} if the weights are perfect, but the map remains {\it unbiased} as long as identical weights are used for the left and right sides of the equation, and the weights are uncorrelated with the sky signal.  
This is the essential part of the method. 
Time-stream filters are by design biased and must be accounted for with simulations.
On the other hand, in our treatment, modes in the map that might cause problems are de-weighted as opposed to being filtered, producing an unbiased solution through careful consideration of the noise structure of the data.  

One further difference remains between these maps and those used in \citet{fowler/etal:2010}.
Here, for every single time-stream sample cut, we fit for its (unknown) amplitude as part of the map solution, as suggested by, for example, \cite{patanchon/etal:2008}.
In addition to cuts from cosmic rays and the like, we explicitly cut the first and last second of each TOD to give the mapping algorithm the freedom to match the TOD beginning and end as smoothly as possible.
Since these data are already highly downweighted by the time-domain taper, the additional amount of data loss is negligible.

\subsection{Map Solution}
With these pre-processing steps, form of the noise, and form of the solution specified, we must then actually solve for the map.
We use the classic Preconditioned Conjugate Gradient algorithm, using the hit-count map as a Jacobi (diagonal) preconditioner for the sky map part of the solution.
To recover the scales around $\ell \simeq 200$ and below, we need a few hundred PCG iterations.
The map making is quite CPU-intensive, with each iteration taking
about a minute on $1600$ 2.53\giga\hertz~Nehalem cores on the Scinet
GPC, or a full run to 1000 iterations taking about a day of wall clock
time, or 5 years of CPU time.  

We note that care must be taken when estimating the noise to make sure it does not bias the map.  
Consider the following thought experiment: Two detectors are scanned across a source, and their weights are estimated from the internal scatters of their time streams.  
In general, noise in the detectors will lead to one of them observing a lower flux from the source.  
If the detector weights are then measured and the source is not removed from the time streams, the detector which
happened to measure the lower flux will usually have a lower variance in its data, and will therefore on average receive more weight than the other detector.  
When the detectors are combined, the more heavily weighted low-flux measurement will lead to the source flux being systematically biased low.
The magnitude of the bias is set by the SNR in the individual time streams, not the final SNR.
The bias can be mitigated if an estimate of the signal is removed from the time streams before noise estimation.  
Since the CMB is highly subdominant to the noise in ACT time streams, this effect is small, but simulations show that it is not negligible.  
So, first we make an initial map using the full dataset and the noise measured directly from the time streams, then we subtract that map from the time streams to reduce the sky signal in them before estimating the noise for a second time, and finally we use this improved estimate of the noise to make the final maps.

\begin{figure}[htbp]
\begin{center}
  \includegraphics[width = 85 mm]{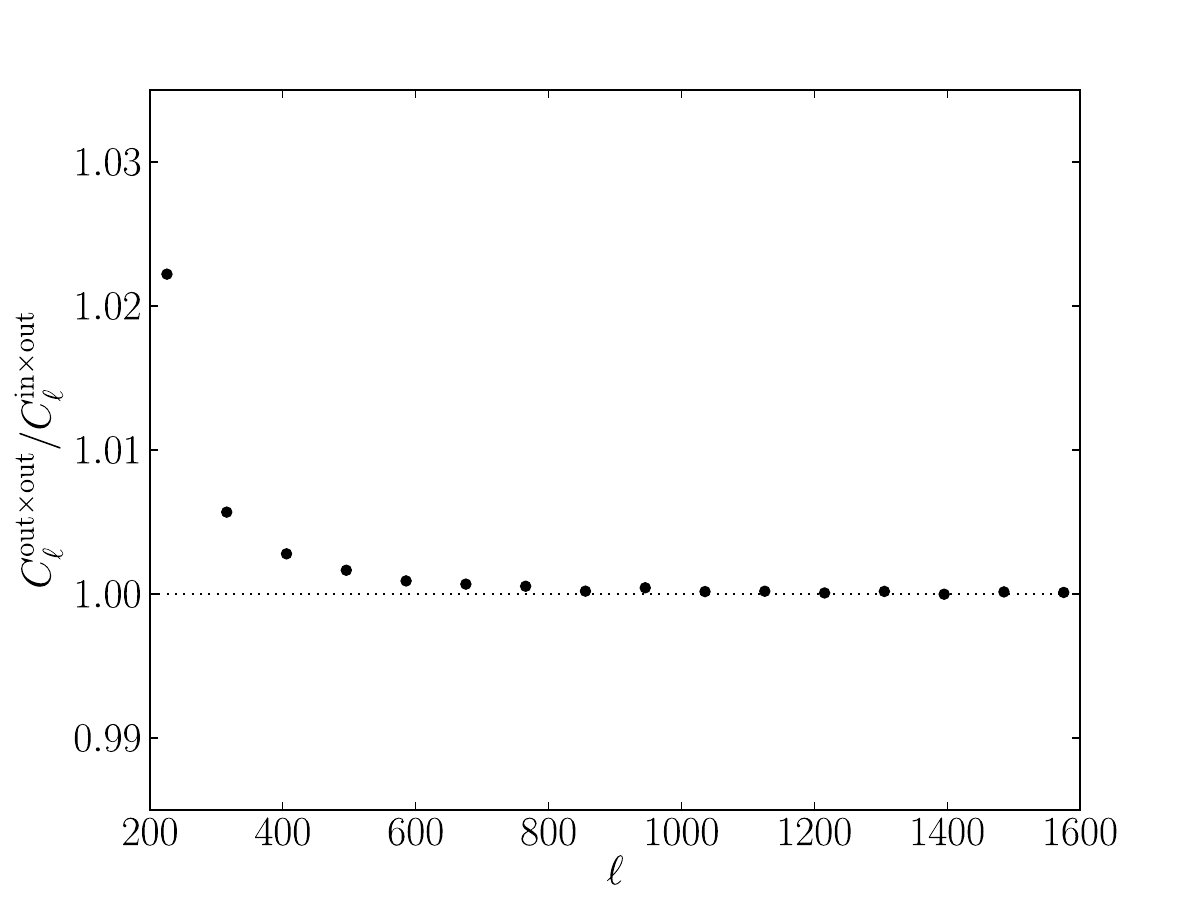}
\caption{Transfer function of the map making process. 
A simulated map (in) was added to the time streams and mapped together with the real data (sim-inject map). 
Then the real data map (previously computed) was subtracted from the sim-inject map to produce the output map (out).
The transfer function was computed as the power spectrum of the output map divided by the cross-spectrum of the output map and the simulated map.
The result is essentially unity for all scales above our noise-estimation input map filter scale of 300 (see text), with a slight boosting of large scales that reaches 2\% at multipoles of 200.
}
\label{trf_fnc}
\end{center}
\end{figure}

ACT must take a bit more care than all-sky experiments because of the large change in sensitivity across the map.
In particular, near the map edges, only a small amount of data contributes to the map, and so removing the raw map would lead to an artificial reduction in the noise of the data near the edges, which would lead to an artificial increase in their weight.  
To prevent this from happening, the starter maps used are first filtered and apodized.  
The apodization is generated by first rescaling the hit count map to the 0-1.0 range, then setting all values higher than a threshold to unity, and finally smoothing the resulting weight map with a Gaussian window. 
Furthermore, from the starter data map we filter out scales larger than $\ell=300$ where atmospheric noise is very large, and scales smaller than $\ell=3000$ where the map is highly noise-dominated and simulations show that the bias effect is truly negligible.  
Finally, we generate the starter map by multiplying the the filtered data map with the apodization window. 
To test this procedure, we carry out the two-step procedure on both the real data, and the real data with a simulated map injected into it.
We compare the difference of these two sets of maps to the original simulation, and show the resulting transfer function in Figure \ref{trf_fnc}.  
At the low-pass scale of $\ell=300$, the maps are unbiased to 0.5\%, and rapidly become unbiased to better than a part in $10^3$ on smaller scales.

\subsection{Map Analysis and Results}
The resulting maps cover and area of 845.6\,deg$^2$ on the sky, ranging between \hmin{20}{43} and \hmin{7}{53} in right ascension and \mbox{$-48$\fdeg1} and \mbox{$-57$\fdeg2} in declination.
A total of five maps were made: one using the full dataset, plus four partial versions using independent subsets of the data for noise estimation purposes.
Each map is accompanied with a hit map with the number of data samples that fell on every pixel of the corresponding map.

These maps are an overall improvement over those used in \citet{fowler/etal:2010, marriage/etal:2011a, marriage/etal:2011b, dunkley/etal:2011, das/etal:2011, hajian/etal:2011, sehgal/etal:2011}, marked particularly by lower noise on angular scales between $500<\ell<2000$, reducing the variance by up to a factor of 2 or 3.  
However, the original maps were already nearly sample-limited on these scales, so errors on the power spectra are only modestly reduced.  
These changes are due to slightly different data selection cuts, improvement in the noise treatment during the mapping process and improved per-detector calibration.

\begin{figure*}
\begin{center}
  \includegraphics[trim = 20 0 0 0, width = 180mm]{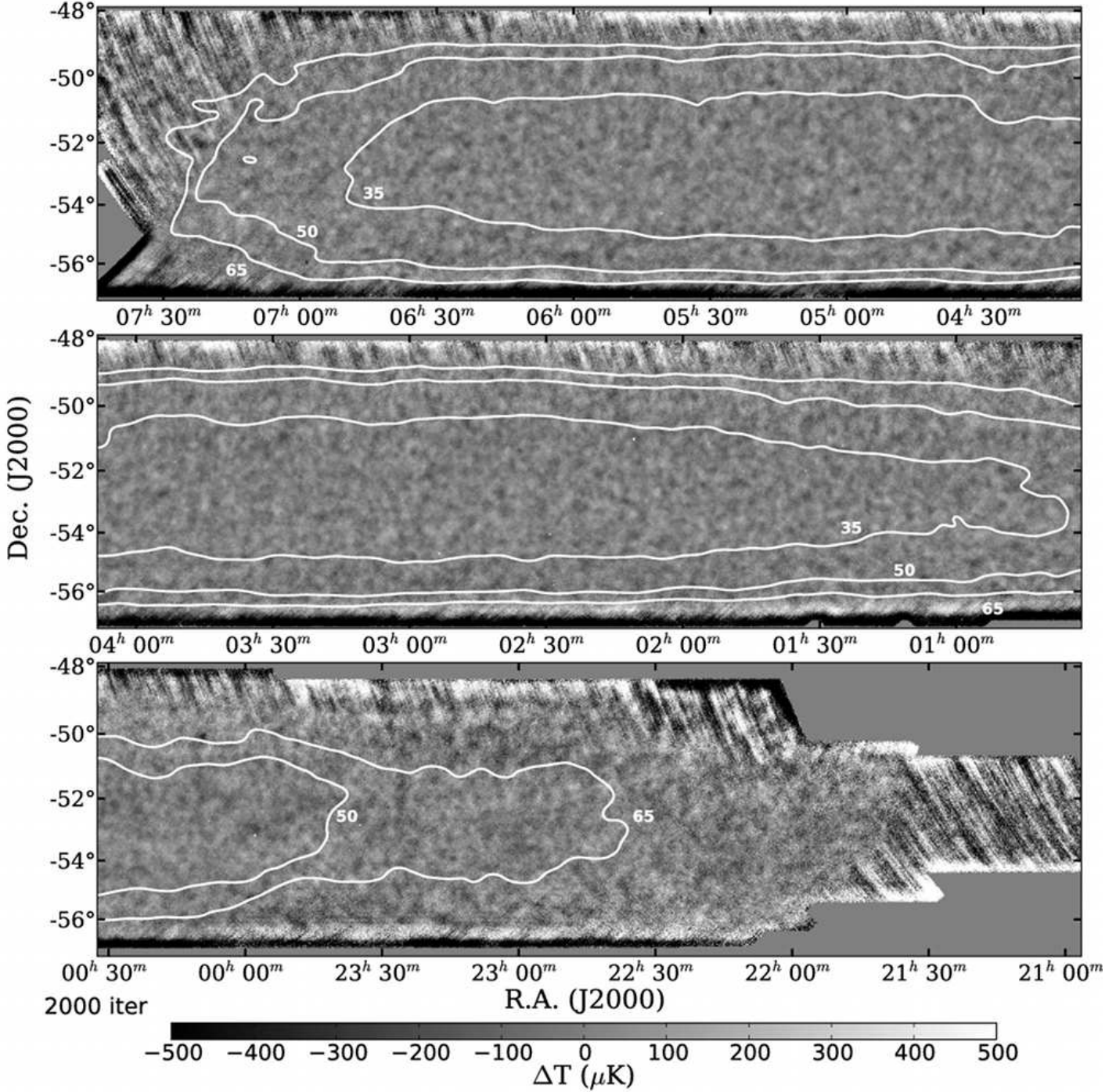}
\caption{Final map of the southern region observed at \arone~in 2008 obtained after 2000 PCG iterations. 
The contours in white join regions with the same sensitivity, namely 35, 50 and 65\,$\mu$\kelvin-arcminute, with increasing sensitivity towards the center of the maps.
The total area enclosed by the contours is 220\,deg$^2$, 420\,deg$^2$ and 530\,deg$^2$ respectively.
The region between $18^h54^m$ and $21^h15^m$ in right ascension is not shown because it was sparsely observed and has low sensitivity.
Most of the analysis has been done in the deeper region between roughly $0^h$ and $7^h$ in R.A..
The map has been high-pass filtered for clarity, depressing modes larger than $\ell = 300$.
The CMB anisotropies can be seen by eye.
There is some evidence of large scale systematic noise, especially near the edges, observed as long horizontal features, which are mostly related to scan-synchronous systematics (such as ground spillover). 
Despite the size of the plot, it is still possible to see some bright point sources and SZ clusters.
}
\label{map}
\end{center}
\end{figure*}

Figure \ref{map} shows the final 2008 southern map, overlaid by contours of iso-sensitivity (35, 50 and 65\,$\mu$\kelvin-arcminute), as different regions of the map have different integration times.
These sensitivities were computed using the number of samples per unit area and the noise equivalent temperature given in Table \ref{sensitivities}.
We run for 2000 PCG iterations on this map to ensure that it is truly converged, though as mentioned we find that all science scales are converged in far fewer PCG iterations.

\begin{figure*}
\begin{center}
  \includegraphics[trim = 100 0 100 0, clip = true, width = 180 mm]{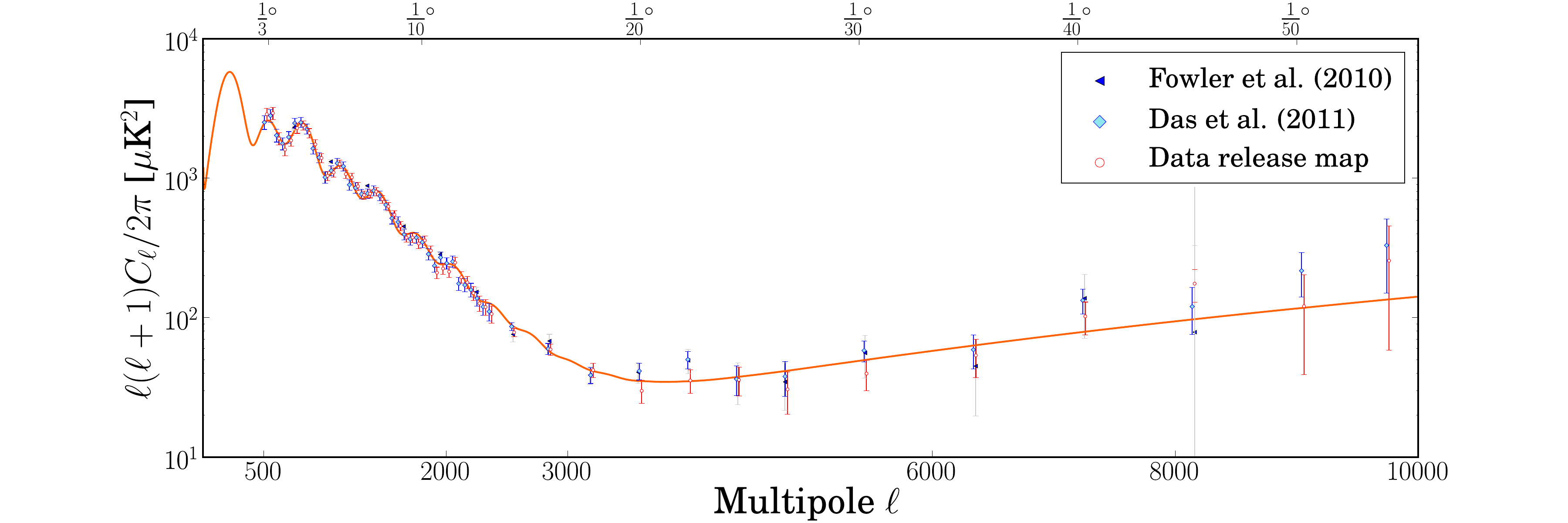}
\caption{Power spectrum of the \arone~map obtained after 2000 PCG iterations compared to the power spectra in \cite{fowler/etal:2010} and \cite{das/etal:2011}.
The thick orange curve shows the best-fit model including the CMB secondaries and point-source contribution taken from \cite{dunkley/etal:2011}.
As revealed by the error bars, the power spectrum derived from this map represents an improvement from previous maps.
}
\label{spectrum}
\end{center}
\end{figure*}

Figure \ref{spectrum} shows the power spectrum computed for this map, compared to previous versions already published in \cite{fowler/etal:2010} and \cite{das/etal:2011}.
The uncertainty in the spectrum reveals an improvement in the quality of the map.

The parameters obtained using a fit to the spectrum released with this paper are slightly shifted with respect to the values presented in \cite{dunkley/etal:2011}.
The largest shifts in the primary parameters were for $n_s$ and $\Omega_b\,h^2$ at $\Delta\sigma = 0.3$ and $\Theta_A$ with $\Delta\sigma = 0.7$, although the change in the inferred dark energy density changed by a negligible $0.03\,\sigma$. 
The other parameters changed by less than $0.07\,\sigma$. 
The secondary parameters were affected by the new mapping procedure, decreasing the Poisson point source level from $13.7\,\mu$K$^2$ to $12.5\,\mu$K$^2$.  
The upper 95\% confidence limits of the correlated point source amplitude and SZ amplitude are instead unchanged. 

Also, a cross-correlation analysis to the BLAST maps (\url{http://blastexperiment.info/}) show correlations detected at a 25$\,\sigma$ level, implying a detection of emission by radio and dusty star-forming galaxies (DSFGs) at high $\ell$, as described in \cite{hajian/etal:2012}.

Finally, the flux of sources as estimated in the data release is approximately 5\% higher than those reported in \cite{marriage/etal:2011a}, due to the new source treatment in the map making and a change in the beam solid angle.
The best representation of the beam shape, as well as the window function for power spectrum analysis, are released together with the maps.

\section{Conclusions}
\label{conclusions}

We have fully characterized the data from the 2008 season of observations from ACT, including data selection, calibration, pointing, random and systematic noise and atmosphere contamination.

Observations done in 2008 yielded a total of nearly 1260\,\hour~of CMB survey data at \arone, distributed between two observation stripes centered at declinations near $0\degree$ and $-53\degree$, with approximately 850 and 280 square degrees respectively.
After data selection, the observing time was reduced to nearly 816\,\hour.
Out of a potential 1024 live detectors, the number of effective detectors was on average 593.
Combining observing time and detector performance, it yields a system efficiency of 38\%.
Including the ratio of observing time over calendar time, the overall system efficiency was 15\%.

The uncorrelated noise in the data, which excludes the atmospheric fluctuations, is dominated by detector noise.
Between 5\,\hertz~and 20\,\hertz, we found a total NET of $32\,\mu\kelvin\sqrt{\second}$ at \arone.
In other words, given 1\,\hour~of observation per square degree, or 1\,\second~of integration per square arcminute, the noise should be about $32\,\mu\kelvin$ arcminutes.

\begin{figure*}
\begin{center}
  \includegraphics[width = 180mm]{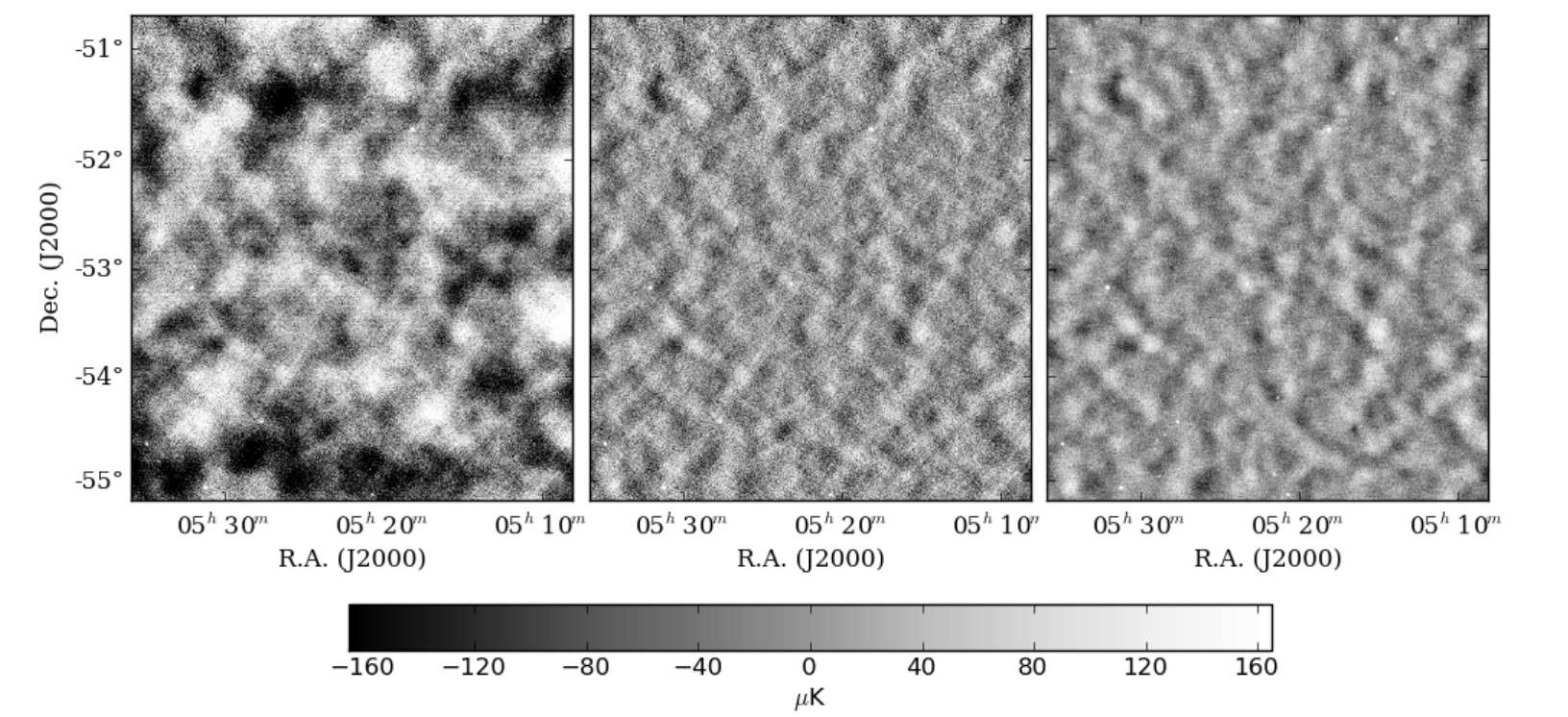}
\caption{
Side by side comparison between the ACT map (season 2008) and the SPT map for the same region of the sky. The left panel shows the ACT map high-pass filtered with a $\cos^2\ell$-like filter that goes from 0 to 1 for $100<\ell<300$, and the center and right panels show the ACT and SPT maps respectively under the same high-pass filter used in the SPT data release \citep{schaffer/etal:2011}. Agreement between the CMB features in the two maps is clear by eye.
}
\label{map_SPT}
\end{center}
\end{figure*}

The correlated noise is dominated at low frequencies by the atmosphere and thermal drift, while at higher frequencies electromagnetic pickup and mechanical vibrations are the largest sources of correlated noise.
The former are characterized in frequency by negative power laws, equaling the detector white noise spectra at approximately 2\,\hertz.
The latter show up mostly as narrow-band signals emerging from the detector noise. 
They can all be modeled as correlated modes added to each detector TOD.
Once identified, they can be de-weighted in the map solution, minimizing the errors in the map, while keeping the solution unbiased.

The data processing steps for map making can be summarized as follows:
The median signal level of each detector is removed together with a single array-wide slope across the 15-minute file, and the inverse anti-alias filter and time constant deconvolution are applied (see \S\ref{random_noise} and \S\ref{time-constants}).
Then a pre-calculated data selection, calibration and pointing solution are applied (see \S\ref{data_selection}, \S\ref{calibration} and \S\ref{pointing}).
After this, 7 pre-computed dark modes are de-projected from each detector TOD (see \S\ref{map-making} and Appendix \ref{mode_removal}).
Before map making, the expected signal from the bright point sources is subtracted from the time streams.
Then the maps are made by minimizing $\chi^2$ over a noise model which includes correlated modes, cuts (missing time streams) and frequency dependence, as described in \S\ref{map-making}.  
This is done twice, with an estimate of the map removed from the data before the noise estimations are repeated in the second mapping pass.
Finally, the flux from point sources previously removed is re-added to the maps.
The resulting maps cover 845.6\,deg$^2$ on the sky and are consistent with WMAP at angular scales measured in common.

The same area of the sky covered by these maps was observed by the South Pole Telescope (SPT) team, who have recently made their data public \citep{schaffer/etal:2011}.
Figure \ref{map_SPT} provides a side by side comparison between the ACT map and the SPT map for the same region of the sky and using the same filtering used in the SPT release.
The correlation between CMB features in the two maps is clear to the eye, supporting the excellent quality of both measurements.
It is also clear that the ACT map is noisier than the SPT map, which will improve in future map releases when data from the following observing seasons (from years 2009 and 2010) are included.

Sky maps for the 148 GHz ACT southern observations from 2008, described in this paper, are available through NASA's Legacy Archive for Microwave Background Data Analysis (LAMBDA), where a variety of ACT analysis software, data products,  and model templates are also available.
Future data releases will include ACT observations at higher frequencies and subsequent observing seasons,  as well as sky coverage in ACT's equatorial stripe which overlaps numerous other observing programs.

\acknowledgements{
This work was supported by the U.S. National Science Foundation through awards AST-0408698 and AST-0965625 for the ACT project, and PHY-0855887 and PHY-1214379.
Funding was also provided by Princeton University, the University of Pennsylvania, and a Canada Foundation for Innovation (CFI) award to UBC.
ACT operates in the Parque Astron\'omico Atacama in northern Chile under the auspices of the Comisi\'on Nacional de Investigaci\'on Cient\'ifica y Tecnol\'ogica de Chile (CONICYT). 
Computations were performed on the GPC supercomputer at the SciNet HPC Consortium.
SciNet is funded by the CFI under the auspices of Compute Canada, the Government of Ontario, the Ontario Research Fund -- Research Excellence; and the University of Toronto.
We specially wish to thank Astro-Norte, Masao Uehara, Felipe Rojas, Patricio Gallardo, Omelan Strysak, Bill Page, Katerina Visnjic, Ben Schmidt, David Faber and Benjamin Walter.
R. D. received additional support from a CONICYT scholarship, from MECESUP, from Fundaci\'on Andes, from FONDECYT-11100147, from Centro de Astrof\'isica y Tecnolog\'ias Afines CATA del Proyecto Financiamiento Basal PFB06 and from Centro de Astrof\'isica FONDAP 15010003.
N. S. is supported by the U.S. Department of Energy contract to SLAC no. DE-AC3-76SF00515 and by the NSF under Award No. 1102762.
E. R. S. acknowledges support by NSF Physics Frontier Center grant PHY-0114422 to the Kavli Institute of Cosmological Physics.
A. K. has been supported by NSF-AST-0807790 for work on ACT. 
R. H. acknowledges funding from the Rhodes Trust and Christ Church.
We are grateful for the assistance we received at various times from the ALMA, APEX, ASTE, CBI/QUIET, and NANTEN2 groups.
}

\appendix

\section{Mode Selection and Removal}
\label{mode_removal}

Contaminant signals, like the atmosphere and systematic signals, produce correlations between detector TODs.
Removing this ``correlated'' noise is important for both map making and data characterization, so we devote this appendix to explain how this is done in more detail.

A contaminant signal can be modeled as a time stream superimposed on the TODs from a set of individual detectors (common mode).
We call this time stream a ``correlated mode.''
One way to estimate it is using an appropriate linear combination of detector TODs. 
For instance, the common mode, defined as a linear combination with equal weight of all detector TODs, is a good estimator of the atmosphere signal.
Organizing all the detector TODs into columns of an $n_{sample}$ by $n_{detector}$ matrix $\mathbf{A}$, the correlated mode $\mathbf{\hat{m}}$ can be expressed as 

\begin{equation}
\mathbf{\hat{m}} = \mathbf{A}\mathbf{\hat{v}}s^{-1},
\label{linear_comb}
\end{equation}

\noindent where $\mathbf{\hat{v}}$ is a vector of unit magnitude representing the $n_{detector}$ coefficients of the linear combination, and $s$ normalizes the mode so it also has unit magnitude.
We henceforth define the correlated modes such that they are always normalized.

Once a correlated mode has been identified, it can be fitted and subtracted from the data:

\begin{equation}
\label{de-projection}
\mathbf{A}' = \left(\mathbf{I} - \mathbf{\hat{m}\hat{m}}^T\right) \mathbf{A} =
              \mathbf{A}\left(\mathbf{I} - 
              \mathbf{\hat{v}\hat{v}}^T\mathbf{A}^T\mathbf{A}\,s^{-2}\right).
\end{equation}

We can also construct correlated modes $\bf\hat m$ in other ways besides the linear combinations in Eq. \ref{linear_comb} (for example, from a thermometry time stream), and in this case the first equality in Eq. \ref{de-projection} must be used directly, rather than the second equality. 

One can identify important correlated modes using the singular value decomposition (SVD) of the matrix $\mathbf{A}$:
\begin{equation}
\mat{A} = \mat{U}\,\mat{S}\,\mat{V}^T,
\end{equation}
where the columns of $\mat{U}$ can be identified as normalized correlated modes $\mathbf{\hat{m}}$, $\mat{S}$ is a diagonal matrix containing the singular values of $\mat{A}$ (identified as the normalization factor, $s$, in Equation \ref{linear_comb}), and the columns of $\mat{V}$ are the eigenvectors of the covariance matrix $\mat{A}^T\mat{A}$ (identified as vector $\mathbf{\hat{v}}$ in Equation \ref{linear_comb}).
The SVD mode selection can be tailored to specific signals by first finding the common mode from subsets of detector TODs, and then applying the previous method to find the strongest correlated modes out of this reduced set of modes.
For example, to find modes that are more likely to correlate in rows, we first find the common mode from all detector TODs in each row (32 row common modes in total), and stack them together as columns of a matrix $\mathbf{A}_{row}$.
The strongest modes are then readily found using SVD.
This method is useful to identify row- and column-correlated modes, and sub-array-scale modes from the atmosphere. 
For the latter, we divide the array into 16 square blocks, finding the common mode in each one before applying the SVD. 
We call the set of modes found this way a ``multi-common mode.''


An important consideration when using modes obtained from linear combinations of live detectors is that they are also correlated to the sky signal.
For example, naively removing the common mode effectively filters the sky, depressing scales larger than the array size ($24\arcmin$ or $\ell\lesssim450$).
Naturally, the effect is stronger for the multi-common mode, which filters scales larger than a fourth of the array ($6\arcmin$ or $\ell\lesssim1800$).
For this reason, the multi-common mode is used to calculate the drift error, but it cannot be naively removed for making maps, nor can any other mode obtained from live detectors.

In Section~\ref{map-making} the appropriate way of projecting modes out of the data is discussed in the context of map making.

\subsubsection{Dark Mode Removal}

The dark detectors share the same readout as the live detectors, so they can capture systematics like thermal drift, electromagnetic pickup and magnetic pickup.
Moreover, as they are not coupled to the optical signal, correlated modes obtained from dark detectors (dark modes) can be safely removed as a preprocessing step before making maps.
For example, the thermal drift is well represented by the low-pass-filtered common mode of the dark detectors.

For row- and column-correlated electromagnetic signals, the ability to identify correlated modes depends on whether the desired row or column of the array contains a working dark detector.
The array design uses one dark detector per column, always in the same row. 
This is ideal for identifying column-correlated modes, but not for row-correlated modes.
For the latter we must rely on the broken live detectors with a properly working readout circuit, as described in \S\ref{data_selection}.
Nevertheless, the row- and column-correlated modes can still be identified by using SVD analysis over all dark detectors.
Note that it is useful to first remove the slow thermal drift from the dark detectors before trying to find these other higher frequency signals. 
Figure \ref{sync_Mcm_corr} shows an example of correlation matrices before and after having removed the dark modes.
Note that both column and row correlations are significantly suppressed after the removal.

\bibliography{act}
\bibliographystyle{apj}

\end{document}